\documentclass[prb, preprint, unsortedaddress, showpacs, nofootinbib, eqsecnum]{revtex4}

\usepackage{amsfonts, amsbsy, amssymb, amsmath, graphicx, color,  float}

\newtheorem{theorem}{Theorem}

\restylefloat{figure}
\newcommand{\rmd}{{\rm d}}

\newcommand{\calM}{\mathcal{M}}

\newcommand{\calP}{\mathcal{P}}

\newcommand{\pb}{\bar{p}}

\newcommand{\sbar}{\overline{s}}

\newcommand{\Int}[3]{\int_{#2}^{#3}\rmd {#1} \;}

\newcommand{\Ab}{\bar{A}}

\newcommand{\Hb}{\bar{H}}

\newcommand{\bz}{\boldsymbol{z}}

\newcommand{\bbP}{\mathbb{P}}

\begin{document}


\title{Isomerization dynamics of a buckled nanobeam}



\author{Peter Collins}
\affiliation{School of Mathematics  \\
University of Bristol\\Bristol BS8 1TW\\United Kingdom}

\author{Gregory S. Ezra}
\email[]{gse1@cornell.edu}
\affiliation{Department of Chemistry and Chemical Biology\\
Baker Laboratory\\
Cornell University\\
Ithaca, NY 14853\\USA}

\author{Stephen Wiggins}
\email[]{stephen.wiggins@mac.com}
\affiliation{School of Mathematics \\
University of Bristol\\Bristol BS8 1TW\\United Kingdom}


\date{\today}

\begin{abstract}
We analyze the dynamics of a model of a nanobeam under compression. 
The model is a two mode truncation of the Euler-Bernoulli beam equation subject 
to compressive stress applied at both ends. We consider parameter regimes where the 
first mode is unstable and the second mode can be either stable or unstable, 
and the remaining modes (neglected) are always stable. 
Material parameters used correspond to a silicon nanobeam. 
The two mode model Hamiltonian is
the sum of a (diagonal) kinetic energy term and a potential energy term. 
The form of the potential energy function suggests an analogy with isomerisation reactions in chemistry,
where `isomerisation' here corresponds to a transition between two stable beam configurations.  
We therefore study the dynamics of the buckled beam using the conceptual framework 
established for the theory of isomerisation reactions.  
When the second mode is stable the potential energy surface has an index one saddle 
and when the second mode is unstable the potential energy surface has an index two 
saddle and two index one saddles. 
Symmetry of the system allows us to readily construct a phase space dividing surface between the two 
``isomers'' (buckled states); we rigorously prove that, in a specific energy range,
it is a normally hyperbolic 
invariant manifold. The energy range is sufficiently wide that we can 
treat the effects of the index one and index two saddles on the isomerisation 
dynamics in a unified fashion. We have computed reactive fluxes, mean gap times and 
reactant phase space volumes for three stress values at several different energies. 
In all cases the phase space volume swept out by isomerizing trajectories is considerably 
less than the reactant density of states, proving that the dynamics is highly nonergodic.  
The associated gap time distributions consist  of one or more `pulses' of trajectories.
Computation of the reactive flux correlation function 
shows no sign of a plateau region; rather, 
the flux exhibits oscillatory decay, indicating that, for the 2-mode model in the physical 
regime considered, a rate constant for isomerization does not exist.
\end{abstract}

\pacs{05.45.-a, 62.25.Fg, 62.25.-g, 82.20.-w, 82.20.Pm, 82.20.Sb, 82.20.Db, 83.10.Ff, 89.90.+n}

\maketitle


\section{Introduction}
\label{sec:intro}

There is currently much 
interest in the mechanical properties of nanoscale objects such as rods and cantilevers 
\cite{Craighead,Roukes,Blencowe,Lautrup}.
For example, changes in resonant frequencies with mass loading
can enable sensitive detection of molecular species with given mass.
The possible manifestation of quantum effects are also of great interest
(see ref.\ \onlinecite{poot12} for a recent review).
The effects of mechanical stress on nanostructures, such as the
buckling of nanobeams, has been studied both experimentally and theoretically 
\cite{lawrence02,carr05,lawrence06}. This latter work is directly related to the work in this paper.

The system studied in the present paper is a nanobeam subject to compressive stress.
According to standard continuum mechanics, the 
beam will buckle as the magnitude of the associated compressive strain increases.
More specifically, following previous work \cite{chakra07,chakra09,chakra11}, 
we consider a regime in which the dynamics of the buckled beam
is usefully described by a 2-mode model obtained by truncation of the
full dynamics.  In the regime considered, one of the modes is
always unstable, while the second mode can be either stable or unstable, and all remaining modes
(which are not considered explicitly) are stable.

We derive a 2 degree-of-freedom (DoF) Hamiltonian describing the 2-mode dynamics.
The Hamiltonian describes a bistable (reactive) mode coupled to a transverse
degree of freedom; the dynamical system thereby obtained has precisely the form of 
simple model potentials
that have been used to describe isomerization reactions in chemistry 
\cite{Robinson72,Forst73,Chandler78,DeLeon81,Berne82,Davis86,Gray87,Chandler87,Baer96}.  However,
for the nanobeam problem, the simple form of the potential is a rigorous
consequence of the 2-mode truncation of the dynamics rather than an approximation to
an unknown molecular potential energy function.

We can therefore investigate the dynamics of the buckled nanobeam using the
conceptual framework established for the theory of isomerization
reactions.  
A number of approaches to the study of isomerization dynamics have been
developed in the chemical literature (see, for example, refs 
\onlinecite{Berne82,Davis86,Gray87,Baer96,Ezra09a}).
Some basic relevant questions are the following: can `isomerization' of the nanobeam
be characterized by a \emph{rate}? If so, can the rate
coefficient be predicted using standard (so-called statistical)
theories \cite{Berne82,Davis86,Baer96}, such as transition state theory 
\cite{Robinson72,Forst73,Pechukas81,Baer96,Waalkens08}? 
If not, what are the dynamical properties of the system that lead to
non-statistical reaction dynamics?

Chakraborty et al.\ have previously 
applied harmonic transition state theory (TST) to predict
isomerization rates for various nanobeams under compressive stress
\cite{chakra07,chakra09,chakra11}.
The rates obtained using basic harmonic TST are proportional to the
inverse of the curvature of the potential transverse to the reaction coordinate,
and so diverge at the strain value for which the second mode
passes from being stable to unstable (and the associated saddle point passes from being
index 1 to index 2 \cite{Ezra09,Collins11,Haller10,Haller11}). 
Chakraborty has also applied a quantum version of harmonic TST to
the nanobeam isomerization kinetics \cite{chakra07,chakra09,chakra11}.

In the present paper we examine the dynamics of a 
compressed buckled Silicon (Si) nanobeam from the
perspective of reaction rate theory.  The potential function which emerges
from a standard modal analysis of the transverse beam displacements has
a high degree of symmetry, and the appropriate dividing surface for computation
of reactive flux is then in fact completely determined by symmetry.
Moreover, we are able to rigorously prove  that the dividing surface is a normally hyperbolic invariant manifold in a specific energy range. The energy range is sufficient for treating the effects of the index one and index two saddles on the isomerisation dynamics in a unified fashion. 

We compute both the distribution of gap times and the reactive flux 
at a number of energies in a physically relevant range.
Our results show that, in this regime for the Si nanobeam
considered:   (i) the isomerization dynamics is extremely regular and nonergodic, and 
(ii) a rate constant for isomerization does not exist.
Rather than exhibiting a rapid drop to a `plateau' value followed
by slow exponential decay \cite{Chandler78,Berne82,Chandler87}, 
the reactive flux shows damped oscillatory decay.
The system  considered is in a regime ($T \gtrsim 100$ K) 
where quantum effects are likely to be negligible.

The structure of the paper is as follows.  In Sec.\ \ref{sec:eom}
we derive the equations of motion for the nanobeam.
In Sec.\ \ref{sec:2mode} we discuss in further detail the
two degree-of-freedom Hamiltonian system obtained for the
2-mode truncation of the beam dynamics.
Particular attention is given to the phase space geometry, 
specifically, the existence of a normally hyperbolic invariant manifold (NHIM) \cite{Wiggins94} 
in the system phase space.
In Sec.\ \ref{sec:volumes} we review some concepts from
reaction rate theory: phase space dividing surfaces and volumes, gap times and
reactive fluxes.
In Sec.\ \ref{sec:param} we discuss the physical parameter values and energy scales
appropriate to our calculations.
Sec.\ \ref{sec:results} presents the results of our numerical calculations, and
Sec.\ \ref{sec:conc} concludes.
In Appendix \ref{sec:nhim_exist} we apply the concept of 
exponential dichotomies \cite{coppel} to provide a proof of the existence of a NHIM 
in the phase space of the truncated nanobeam problem.

\newpage
\section{Equations of Motion for the Nanobeam}
\label{sec:eom}

In this section we derive a 2-mode (classical) model for an Euler-Bernoulli beam subject 
to compressive stress applied at both ends.
The  derivation of the Euler-Bernoulli equations can be 
found in many textbooks on continuum mechanics (e.g. ref.\ \onlinecite{Lautrup}).
Our derivation and notation closely follows 
that of refs \onlinecite{chakra07, chakra09, chakra11} 
(see also refs \onlinecite{lawrence02, carr05, lawrence06}).
A useful discussion of the concept of stress in a quantum mechanical system 
is given in ref.\ \onlinecite{Maranganti10}.

We consider an  Euler-Bernoulli model of a beam of length $L$ having width $w$ and thickness $d$, where $L \gg w>d$.
The requirement $w>d$ allows us to assume that
transverse displacements, $y(x, t)$, occur only in the $d$ direction.
The  linear modulus $F$ (dimensions of energy per unit length)  is
related to the elastic modulus $Q$ by $F=Qwd$.
For a beam of rectangular cross-section the bending moment is given by $\kappa=\frac{d^2}{12}$ and
$\mu = \frac{m}{L}$ denotes the mass per unit length.  The length of the uncompressed rod is denoted  by $L_0$.

Constant compressive stress is applied to both ends of the beam, 
reducing the horizontal distance between the two endpoints of
the beam to $L < L_0$.   The strain $\epsilon$ is
$\epsilon = \frac{L-L_0}{L_0}$, and is negative for compression.  The compression causes a contribution to the potential energy of the beam due
to bending in the $d$ direction (the first term in \eqref{eq:PE_1}) and elasticity (the last three terms in \eqref{eq:PE_1}),
where the potential energy has the form:
\begin{equation}
V[y(x, t)] = \frac{1}{2} \int_{0}^{L} dx \left( F \kappa^2 (y^{\prime \prime})^2 + F \epsilon (y^\prime)^2 \right) + \frac{F}{8L_0} \left( \int_{0}^{L}dx
({y^{\prime}})^2 \right)^2 + \frac{F L_0}{2} \epsilon^2 \,.
\label{eq:PE_1}
\end{equation}
The kinetic energy is:
\begin{equation}
T[\dot{y}(x, t)] = \frac{1}{2} \int_{0}^{L} dx \, \mu \dot{y}^2
\label{eq:KE_1}
\end{equation}
Forming the Lagrangian in the usual manner:
\begin{equation}
L[y(x,t), \dot{y}(x, t)] = T[\dot{y}(x, t)]  - V[y(x, t)]
\label{eq:Lag}
\end{equation}
Lagrange's equations of motion are given by:
\begin{equation}
\frac{d}{dt} \frac{\delta L}{\delta \dot{y}}- \frac{\delta L}{\delta y} =0.
\label{eq:LE}
\end{equation}
Using \eqref{eq:PE_1} and \eqref{eq:KE_1}, we have:
\begin{subequations}
\label{eq:var_deriv}
\begin{align}
\frac{ \delta T[\dot{y}(x, t)] }{\delta \dot{y}(x, t)]}  & =     \mu \dot{y}, \\
\frac{ \delta V[y(x, t)]}{\delta y(x, t)]}  & =  
F \kappa^2 y^{[4]} - \left[  F \epsilon y^{\prime \prime} + 
\frac{F}{2 L_0} \left( \int_{0}^{L} dx (y^{\prime} )^2
\right) \, y^{\prime \prime}\right].
\end{align}
\end{subequations}

Using these expressions, together with \eqref{eq:Lag} and  \eqref{eq:LE},  gives the equations of motion:
\begin{equation}
\mu \ddot{y} =  -F \kappa^2 y^{[4]}  +\left[  F \epsilon y^{\prime \prime} + \frac{F}{2 L_0} \left( \int_{0}^{L} dx (y^{\prime} )^2 \right) \, y^{\prime \prime}
\right],
 \label{eq:eom}
\end{equation}
and the boundary conditions are chosen to be:
\begin{subequations}
\label{eq:BC}
\begin{align}
y(0, t) & = y(L, t)=0, \\
y^{\prime \prime} (0, t) & = y^{\prime \prime} (L, t)=0,
\end{align}
\end{subequations}
which are referred to in the literature as {\em hinged} boundary conditions.

One easily sees by inspecting \eqref{eq:eom}  that $y(x, t)=0$ is a solution. Linearizing  \eqref{eq:eom} about this solution  gives:
\begin{equation}
\mu \ddot{y} =  -F \kappa^2 y^{[4]}  +  F \epsilon y^{\prime \prime}.
 \label{eq:eom_lin}
\end{equation}
We seek the normal modes (eigenfunctions)  of these linearized equations by assuming a  solution of the form:
\begin{equation}
y_n (x, t) = y_n (x) e^{i \omega_n t}, \quad n=1, 2, 3, \ldots
\label{eq:ef_form}
\end{equation}
where
\begin{equation}
y_n(x) = \sqrt{\frac{2}{L}} \sin \left[\frac{n \pi x}{L}\right].
\label{eq:ef_sine}
\end{equation}
Note that the $y_n(x)$ satisfy the normalization condition:
\begin{equation}
\int_{0}^{L} y_n (x) y_m (x) dx = \delta_{n,m}.
\label{eq:normal}
\end{equation}
Substituting \eqref{eq:ef_form} into \eqref{eq:eom_lin} gives:
\begin{equation}
\mu \omega_n^2 y_n (x) = F \kappa^2 y_n^{[4]} (x) - F \epsilon y_n^{\prime \prime} (x)
\label{eq:ef_form_2}
\end{equation}
We substitute \eqref{eq:ef_sine} into \eqref{eq:ef_form_2}, and after some algebra we obtain:
\begin{equation}
\omega_n = \omega_0 n \sqrt{n^2 - \frac{\epsilon}{\bar{\epsilon}}}
\label{eq:mode_freq}
\end{equation}
with
\begin{subequations}
\begin{align}
\omega_0 & = \pi^2 \frac{\kappa}{L^2} \sqrt{\frac{F}{\mu}} \\
\label{eq:critical_strain}
\bar{\epsilon} & = -\frac{\kappa^2 \pi^2}{L^2}.
\end{align}
\end{subequations}
Note that the quantity $\bar{\epsilon}$ is approximately equal to the critical value of the
  strain $\epsilon_c$, obtained by solving the implicit equation
  \eqref{eq:critical_strain}, but has a
  weak dependence on $\epsilon$.

From the form of \eqref{eq:ef_form} and \eqref{eq:mode_freq}, we see that the mode $y_n (x, t)$ is
linearly stable (resp.,  unstable) provided $n^2 - \frac{\epsilon}
{\bar{\epsilon}} > 0$ (resp., $n^2 - \frac{\epsilon}{\bar{\epsilon}} < 0$ ).

We will examine the situation where:
\begin{subequations}
\begin{align}
& 1- \frac{\epsilon}{\bar{\epsilon}} < 0 ,  \\
&  4- \frac{\epsilon}{\bar{\epsilon}}  >  0  \quad \mbox{or} \quad 4- \frac{\epsilon}
{\bar{\epsilon}}  <  0 ,  \\
& n^2- \frac{\epsilon}{\bar{\epsilon}}  >  0, \quad n \ge 3.
\end{align}
\end{subequations}
In other words, we will consider the cases where the first mode is always unstable,
the second mode can be either stable or unstable, and modes $n \ge 3$ are all stable.

Assuming that the solution of \eqref{eq:eom} has the form:
\begin{equation}
y(x, t) = \sqrt{\frac{2}{L}} \, \sum_{n=1}^{\infty} A_n(t) \sin \left[\frac{n \pi x}{L}\right],
\label{eq:mode}
\end{equation}
we substitute into \eqref{eq:eom} to obtain  
an infinite set of ordinary differential equations for the time evolution 
of the modal amplitudes, $A_n (t)$.
However, we will simplify the problem by only considering the evolution of the first two modes:
\begin{equation}
y(x, t) = \sqrt{\frac{2}{L}} A_1(t) \sin \left[\frac{\pi x}{L}\right] +   \sqrt{\frac{2}{L}} A_2(t) \sin \left[\frac{2 \pi x}{L}\right].
\label{eq:2mode}
\end{equation}
In this case one obtains a  two-degree-of-freedom system for the evolution of the modal amplitudes $A_1 (t)$ and $A_2 (t)$.
Defining momentum variables
$p_i = \mu \dot{A}_i, \, i=1, 2$, the time evolution of the  amplitudes $A_i$
is described by a two degree-of-freedom Hamiltonian system, with Hamiltonian:
\begin{equation}
H = \frac{p_1^2}{2 \mu} + \frac{p_2^2}{2 \mu} + V(A_1, A_2),
\label{eq:ham1}
\end{equation}
where the potential energy has the form:
\begin{equation}
V(A_1, A_2) = \frac{F \pi^2 (\epsilon - \bar{\epsilon})}{2 L^2} A_1^2 + \frac{2F \pi^2 
(\epsilon - 4\bar{\epsilon})}{ L^2} A_2^2
+ \frac{F \pi^4 }{8 L^4 L_0} (A_1^2 + 4 A_2^2)^2 .
\label{eq:pot1}
\end{equation}

It is natural to ask how well the two-mode truncation described by 
the two degree-of-freedom Hamiltonian system defined by
Hamiltonian \eqref{eq:ham1}  models the
solution of the full partial differential equation describing the Euler-Bernoulli 
beam given in \eqref{eq:eom}.
It is a standard engineering approximation to approximate the
full solution of a partial differential equation by considering a 
truncated modal expansion of eigenfunctions
obtained from the linearized equations about an  equilibrium
state. The reasoning is that the evolution near the equilibrium solution 
is dominated by the evolution of the unstable modes.
In some cases this can be rigorously proven
using center manifold or inertial manifold techniques (see, e.g., refs \onlinecite{carr81, temam97}).
A   seminal example of this approach that played a fundamental role in the
development of  applied dynamical systems theory was the work of 
Holmes, Marsden, and Moon on the dynamics of  a buckled beam subject to  periodic (temporal)
forcing \cite{moon79,moon80,holmes81}.
Initially, a combination of experimental and theoretical work  
showed that the experimentally observed
chaotic behavior was captured  by the
evolution of the one unstable mode, subject to forcing \cite{moon79, moon80}. 
Later, it was rigorously shown \cite{holmes81} that this single mode truncation
captured the dynamics of the  full partial differential equation governing the beam (near the instability).
In this paper we will not be concerned with these issues. Rather,
we take as the starting point of our analysis the  two degree-of-freedom Hamiltonian system governing
the two mode truncation of the Euler-beam equation given in \eqref{eq:eom}.

\newpage
\section{Two-mode truncation: Hamiltonian and phase space geometry}
\label{sec:2mode}

We begin by non-dimensionalizing the two degree-of-freedom Hamiltonian system of eq.\ \eqref{eq:ham1}.
Defining the dimensionless amplitudes:
\begin{subequations}
\label{eq:nondim}
\begin{align}
A_1 & =  \frac{L \sqrt{2 L_0}}{\pi} \bar{A}_1,  \\
A_2 & =  \frac{ L \sqrt{2 L_0}}{\pi} \bar{A}_2
\end{align}
\end{subequations}
and substituting these expressions into the potential function \eqref{eq:pot1} gives:
\begin{subequations}
\label{eq:pot2}
\begin{align}
V(\bar{A}_1, \bar{A}_2) & = F L_0 \left[
(\epsilon - \bar{\epsilon}) \bar{A}_1^2 +  (\epsilon - 4\bar{\epsilon}) 4 \bar{A}_2^2 + 
\frac{1}{2} (\bar{A}_1^2 + 4 \bar{A}_2^2)^2 \right] \\
& \equiv F L_0  \bar{V}(\bar{A}_1, \bar{A}_2).
\end{align}
\end{subequations}
Defining associated momenta $\{ \bar{p}_k\}$ conjugate to the $\{ \bar{A}_k\}$ via 
\begin{subequations}
\label{eq:nondim_p}
\begin{align}
\bar{p}_1 & =  \frac{L \sqrt{2 L_0}}{\pi} \, p_1,  \\
\bar{p}_2 & =  \frac{L \sqrt{2 L_0}}{\pi} \, p_2
\end{align}
\end{subequations}
and substituting into \eqref{eq:ham1} gives the scaled
Hamiltonian:
\begin{equation}
\bar{H} \equiv \frac{H}{F L_0}  = 
\frac{\bar{p}_1^2}{2 \bar{\mu}} + \frac{\bar{p}_2^2}{2 \bar{\mu}} +  \bar{V}(\bar{A}_1, \bar{A}_2)
\label{eq:ham2}
\end{equation}
where
\begin{equation}
\bar{\mu} \equiv \mu \frac{2 F L^2 L_0^2}{\pi^2}
\end{equation}
Rescaling the momenta
\begin{equation}
\bar{p}_i \to \frac{\bar{p}_i}{\bar{\mu}^{1/2}}, \quad i=1,2
\label{eq:mom_dim}
\end{equation}
we obtain the following Hamiltonian:
\begin{equation}
\bar{H} = \frac{\bar{p}_1^2}{2 } + \frac{\bar{p}_2^2}{2 } +  \bar{V}(\bar{A}_1, \bar{A}_2),
\label{eq:ham_dim}
\end{equation}
with
\begin{equation}
\bar{V}(\bar{A}_1, \bar{A}_2 )=  \alpha \bar{A}_1^2 + 4 \beta \bar{A}_2^2 
+ \frac{1}{2} (\bar{A}_1^2 + 4 \bar{A}_2^2)^2 ,
\label{eq:pot_dim}
\end{equation}
and
\begin{subequations}
\label{eq:param}
\begin{align}
\alpha & =  \epsilon - \bar{\epsilon}  \\
\beta & =   \epsilon - 4\bar{\epsilon} \,.
\end{align}
\end{subequations}
The corresponding Hamiltonian equations of motion (for suitably rescaled time) are then:
\begin{subequations}
\label{eq:hameq_1}
\begin{align}
\dot{\Ab}_1 & =  \frac{\partial \Hb}{\partial \pb_1} = \pb_1,  \\
\dot{\pb}_1 & =  - \frac{\partial \Hb}{\partial \Ab_1} = - 2 \Ab_1 \left( \alpha + \Ab_1^2 + 4\Ab_2^2 \right),  \\
\dot{\Ab}_2 & =  \frac{\partial \Hb}{\partial \pb_2} = \pb_2,  \\
\dot{\pb}_2 & =  - \frac{\partial \Hb}{\partial \Ab_2} = - 8 \Ab_2 \left( \beta + \Ab_1^2 + 4 \Ab_2^2 \right).
\end{align}
\end{subequations}

\subsection{Equilibria and their stability}
\label{sec:equil}

From the (algebraically) simple form of Hamilton's equations given in eq.\
\eqref{eq:hameq_1} it is straightforward to compute the equilibria, determine their linearized stability
properties (i.e. compute the eigenvalues of the matrix associated with the linearization of Hamilton's equations about the equilibrium point), and compute the (total)
energy of the equilibrium point. These properties are summarized in Table \ref{table:eqinfo}.

For $\alpha < 0$, $\beta >0$  there are only three equilibrium points.
The origin is an index one saddle
point and the remaining two equilibria have two pairs of purely imaginary eigenvalues (minima of the potential \eqref{eq:pot_dim}). For $\alpha < 0$, $\beta < 0$,
$\vert \alpha \vert > \vert \beta \vert$, the origin is an index two saddle,
phase space points $\left(\bar{A}_1, \bar{p}_1, \bar{A}_2, \bar{p}_2 \right) = \left(\pm \sqrt{-\alpha}, 0,
0, 0 \right)$ have two pairs of purely imaginary eigenvalues and correspond to  minima of the potential \eqref{eq:pot_dim},
and $\left(\bar{A}_1, \bar{p}_1, \bar{A}_2, \bar{p}_2
\right) = \left(0, 0, \pm\frac{\sqrt{-\beta}}{2}, 0 \right)$ are index one saddles.

\subsection{Invariant planes and the existence of a Normally Hyperbolic Invariant Manifold}
\label{sec:nhim}

It can be seen  by inspection of \eqref{eq:hameq_1} that,
if we set $\Ab_2 = \pb_2 =0$  (resp., $\Ab_1 = \pb_1 =0$), then $\dot{\Ab}_2 = \dot{\pb}_2 =0$  (resp.,
$\dot{\Ab}_1= \dot{\pb}_1 =0$).  It then follows that the two planes:
\begin{subequations}
\begin{align}
\Pi_1 & =  \left\{ ( \Ab_1, \pb_1, \Ab_2, \pb_2) \, \vert \,  \Ab_2 = \pb_2 =0 \right\},  \label{eq:pi1}\\
\Pi_2 & =  \left\{ ( \Ab_1, \pb_1, \Ab_2, \pb_2) \, \vert \,  \Ab_1 = \pb_1 =0 \right\} \label{eq:pi2}
\end{align}
\end{subequations}
are each invariant with respect to the dynamics generated by \eqref{eq:hameq_1}. The dynamics on $\Pi_1$ is given by the Hamiltonian system defined by the
Hamiltonian 
\begin{subequations}
\begin{equation}
\Hb_1 \equiv  \frac{\pb_1^2}{2} + \alpha \Ab_1^2 + \frac{1}{2} \Ab_1^4
\end{equation}
and the dynamics on $\Pi_2$ is given by the Hamiltonian system defined by
the Hamiltonian
\begin{equation}
\Hb_2 \equiv  \frac{\pb_2^2}{2} + 4 \beta \Ab_2^2 + 8 \Ab_2^4\, .
\end{equation}
\end{subequations}
Hence, the dynamics on each plane is  integrable.  However, the
dynamics on each plane is {\em not} isoenergetic.  
The three dimensional energy surface intersects a two dimensional plane in the four dimensional phase space in a one
dimensional set, i.e. a trajectory of the one degree-of-freedom Hamiltonian system 
defined by $\Hb_1$ (for intersections with $\Pi_1$) or a trajectory of the one 
degree-of-freedom Hamiltonian system defined by $\Hb_2$ (for intersections with $\Pi_2$).

We  now want to determine conditions under which some portion of $\Pi_2$ is a 
normally hyperbolic invariant manifold
(NHIM).  Roughly speaking, NHIMs have saddle-like stability properties in directions 
transverse to the invariant manifolds \cite{Wiggins94}.
In recent years NHIMs  have been shown
to be a significant phase space structure related to reaction dynamics. 
For example, they play the key role in the construction of a
phase space dividing surface having the
no-recrossing property \cite{wwju,ujpyw} and  minimal flux \cite{WaalkensWiggins04}.
They have also been shown to be central to Thiele's theory \cite{Thiele62}
of reaction dynamics in terms of gap times \cite{Ezra09a}.

The following theorem provides sufficient conditions for the existence of a NHIM  (the proof  is given in Appendix  \ref{sec:nhim_exist}):
\begin{theorem}
Consider $\alpha <0$ and the region on the $\Pi_2$ plane bounded by the curve:
\begin{equation}
 \frac{\pb_2^2}{2} + 4\beta \Ab_2^2 + 8 \Ab_2^4 =E_{\text {max}},
 \label{eq:nhim_bound_1}
\end{equation}
where
 \begin{equation}
 E_{\text {max}} = \frac{\alpha^2}{2} \left(1-2 \frac{\beta}{\alpha} \right).
 \label{eq:nhim_bound_2}
 \end{equation}
Then this region  on $\Pi_2$ is a two-dimensional (non-isoenergetic) normally hyperbolic invariant manifold.
\label{nhimthm}
\end{theorem}

Note that for a given three dimensional energy surface the NHIM is a (one dimensional) trajectory
on $\Pi_2$.

\subsection{The existence of a phase space dividing surface having the  no-recrossing property}
\label{sec:DS}

For the Hamiltonian  \eqref{eq:ham_dim} we now construct a dividing surface in phase space
having the ``no-recrossing''  property. We will describe what this means, as well as the dynamical significance of
the dividing surface, in the  course of our construction.

The codimension one non-isoenergetic surface defined by $\Ab_1=0$ divides the phase space into two regions: one associated with the potential well
whose minimum is $\left(\bar{A}_1, \bar{p}_1, \bar{A}_2, \bar{p}_2 \right) = \left( \sqrt{-\alpha}, 0, 0, 0 \right)$
and the other associated with the potential well
whose minimum is $\left(\bar{A}_1, \bar{p}_1, \bar{A}_2, \bar{p}_2 \right) = \left(- \sqrt{-\alpha}, 0, 0, 0 \right)$.
The dividing surface restricted to a fixed energy
surface $\Hb = E$  is given by:
\begin{equation}
DS(E) = \left\{\left(\bar{A}_1, \bar{p}_1, \bar{A}_2, \bar{p}_2 \right) \, \vert \, \Ab_1=0, \, \bar{H} = \frac{\bar{p}_1^2}{2 } + \frac{\bar{p}_2^2}{2 }
 + 4\beta \bar{A}_2^2 + 8  \bar{A}_2^4=E\right\}
\label{eq:DS}
\end{equation}
This dividing surface has two halves:
\begin{subequations}
\begin{equation}
DS_+(E) = \left\{\left(\bar{A}_1, \bar{p}_1, \bar{A}_2, \bar{p}_2 \right) \, \vert \, \Ab_1=0, \, \bar{H} = \frac{\bar{p}_1^2}{2 } + \frac{\bar{p}_2^2}{2 }
 + 4\beta \bar{A}_2^2 + 8  \bar{A}_2^4=E, \, \bar{p}_1 >0\right\}
\label{eq:DS+}
\end{equation}
and
\begin{equation}
DS_-(E) = \left\{\left(\bar{A}_1, \bar{p}_1, \bar{A}_2, \bar{p}_2 \right) \, \vert \, \Ab_1=0, \, \bar{H} = \frac{\bar{p}_1^2}{2 } + \frac{\bar{p}_2^2}{2 }
 + 4\beta \bar{A}_2^2 + 8 \bar{A}_2^4=E,  \, \bar{p}_1 < 0\right\}.
\label{eq:DS-}
\end{equation}
\end{subequations}
These two halves meet at the NHIM:
\begin{equation}
\mbox{NHIM}(E) = \left\{\left(\bar{A}_1, \bar{p}_1, \bar{A}_2, \bar{p}_2 \right) \, \vert \, \Ab_1=0, \, \bar{H} = \frac{\bar{p}_2^2}{2 }
 +4 \beta \bar{A}_2^2 + 8  \bar{A}_2^4=E,  \, \bar{p}_1 = 0\right\}.
\label{eq:DSint}
\end{equation}
The nature of the NHIM (i.e., the boundary between $DS_+(E)$ and  $DS_-(E)$) depends on both $E$ and $\beta$.
The dynamics on the $\Ab_2-\pb_2$ plane is illustrated  in Fig. \ref{fig:nhim}.

We now argue that $DS_+(E)$ and $DS_-(E)$ are surfaces having the no (local) re-crossing property.
These surfaces are defined by $\Ab_1=0$. Therefore points on
these surfaces leave  if $\dot{\Ab}_1 \neq 0$.
We see from \eqref{eq:hameq_1} that $\dot{\Ab}_1  =  \frac{\partial \Hb}{\partial \pb_1} = \pb_1$.  
Therefore on
$DS_+ (E)$  we have $\dot{\Ab}_1 > 0$ and on $DS_- (E)$  we have $\dot{\Ab}_1 < 0$.  
Trajectories through points on $DS_+(E)$ move towards the region of phase
space associated with the potential well whose minimum is 
$\left(\bar{A}_1, \bar{p}_1, \bar{A}_2, \bar{p}_2 \right) = \left( \sqrt{-\alpha}, 0, 0, 0 \right)$  
and
points on $DS_-(E)$ move towards the region of phase space associated 
with the potential well whose minimum is
 $\left(\bar{A}_1, \bar{p}_1, \bar{A}_2, \bar{p}_2 \right) = \left(- \sqrt{-\alpha}, 0, 0, 0 \right)$.

We denote the directional flux
across these hemispheres by $\phi_{\text{+}} (E)$ and $\phi_{\text{-}} (E)$,
respectively, and note that $\phi_{\text{+}} (E)+ \phi_{\text{-}}(E)=0$.
The magnitude of the flux is
$\vert \phi_{\text{+}} (E)\vert =  \vert \phi_{\text{-}}(E)\vert \equiv \phi (E)$.
The magnitude of the flux and related quantities are central to the theory of
isomerization rates, as discussed in Sec.\ \ref{sec:volumes}.

\newpage

\section{Phase space volumes, gap times, and reactive flux}
\label{sec:volumes}

In this section we briefly review the concepts from classical reaction rate theory that
will be applied to the dynamics of the buckled nanobeam.

Points in the $4$-dimensional system phase space $\calM  = \mathbb{R}^{4}$ are denoted
$\bz \equiv (\pb_1, \pb_2, \Ab_1, \Ab_2)  \equiv (\mathbf{\pb}, \mathbf{\Ab} )\in \calM$.
The system Hamiltonian is $\Hb(\bz)$, and
the $3$ dimensional  energy shell at energy $E$, $\Hb(\bz) = E$, is denoted $\Sigma_E \subset \calM$.
The corresponding microcanonical phase
space density is $\delta(E - \Hb(\bz))$, and the associated density of states
for the complete energy shell at energy $E$ is
\begin{equation}
\rho(E) = \Int{\bz}{\calM}{} \delta(E - \Hb(\bz)).
\end{equation}

The disjoint regions of phase space separated by $\text{DS}(E)$  are denoted $\calM_{\pm}$;
the region of phase space corresponding to the
potential well whose minimum is $\left(\bar{A}_1, \bar{p}_1, \bar{A}_2, \bar{p}_2 \right) = \left( \sqrt{-\alpha}, 0, 0, 0 \right)$ will be denoted by
$\calM_{\text{+}}$,  and that corresponding to the
potential well whose minimum is $\left(\bar{A}_1, \bar{p}_1, \bar{A}_2, \bar{p}_2 \right) = \left(- \sqrt{-\alpha}, 0, 0, 0 \right)$  will be denoted by
$\calM_{\text{-}}$.

The microcanonical density of states for points in region $\calM_{\text{+}}$  is
\begin{equation}
\rho_{\text{+}}(E) = \Int{\bz}{\calM_{\text{+}}}{} \delta(E - H(\bz))
\end{equation}
with a corresponding expression for the density of states $\rho_{\text{-}}(E)$ in $\calM_-$.
Since  the flow is everywhere transverse to $\text{DS}_{\pm}(E)$,
those phase points  in the  region $\calM_{\text{+}}$
that lie on crossing trajectories \cite{DeLeon81,Berne82}
(i.e., those trajectories that cross $\text{DS}_{\pm}(E)$) can be  specified uniquely by coordinates $( \widetilde{p}, \widetilde{A}, \psi)$,
where $(\widetilde{p}, \widetilde{A}) \in \text{DS}_{\text{+}}(E)$ is a point on
$\text{DS}_{\text{+}}(E)$, specified by $2$
coordinates $(\widetilde{p}, \widetilde{A})$, and
$\psi$ is a time variable.
The point $\bz(\widetilde{p}, \widetilde{A}, \psi)$ is reached by propagating the
initial condition $(\widetilde{p}, \widetilde{A}) \in \text{DS}_{\text{+}}(E)$ forward for time $\psi$
\cite{Thiele62,Ezra09a}.
As all initial conditions on $\text{DS}_{\text{+}}(E)$
(apart from a set of trajectories of measure zero lying on stable manifolds)
will leave the region $\calM_{\text{+}}$ in finite time by crossing $\text{DS}_{\text{-}}(E)$, for each
$(\widetilde{p}, \widetilde{A}) \in \text{DS}_{\text{+}}(E)$, we can define the \emph{gap time}
$s = s(\widetilde{p}, \widetilde{A})$, which is the
time it takes for the trajectory to traverse the  region $\calM_{\text{+}}$ before entering the region $\calM_{\text{-}}$.
That is, $\bz(\widetilde{p}, \widetilde{A}, \psi = s(\widetilde{p}, \widetilde{A})) \in \text{DS}_{\text{-}}(E)$.
For the phase point $\bz(\widetilde{p}, \widetilde{A}, \psi)$, we therefore have
$0 \leq \psi \leq s(\widetilde{p}, \widetilde{A})$.

The coordinate transformation $\bz \to (E, \psi, \widetilde{p}, \widetilde{A})$ is canonical
\cite{Arnold78,Thiele62,Binney85,Meyer86}, so that the phase space volume element is
\begin{equation}
\label{coord_1}
\rmd^{4} \bz = \rmd E \, \rmd \psi  \, \rmd \sigma
\end{equation}
with $\rmd \sigma \equiv \rmd \widetilde{p} \, \rmd \widetilde{A}$
an element of $2$ dimensional area on the DS(E).

The magnitude $\phi(E)$ of the flux through dividing surface
$\text{DS}_{\text{+}}(E)$ at energy $E$ is given by
\begin{equation}
\label{flux_1}
\phi(E) = \left\vert\Int{\sigma}{\text{DS}_{\text{+}}(E)}{} \right\vert,
\end{equation}
where the element of area $\rmd \sigma$ is precisely the restriction to DS(E) of the
appropriate flux $2$-form $\omega$ corresponding to the Hamiltonian vector field
associated with $\Hb(\bz)$ \cite{Toller85,Mackay90,Gillilan90,WaalkensWiggins04}.
The reactant phase space volume occupied by points initiated on the dividing surface
with energies between $E$ and $E + \rmd E$ is therefore
\cite{Thiele62,Brumer80,Pollak81,Binney85,Meyer86,WaalkensBurbanksWiggins05,WaalkensBurbanksWiggins05c,Ezra09a}
\begin{subequations}
\label{vol_1}
\begin{align}
\rmd E \Int{\sigma}{\text{DS}_{\text{+}}(E)}{} \Int{\psi}{0}{s}
& = \rmd E \Int{\sigma}{\text{DS}_{\text{+}}(E)}{}  s \\
&= \rmd E \,\, \phi(E) \, \sbar
\end{align}
\end{subequations}where the \emph{mean gap time} $\sbar$ is defined as
\begin{equation}
\sbar = \frac{1}{\phi(E)} \, \Int{\sigma}{\text{DS}_{\text{+}}(E)}{}  s
\end{equation}
and is a function of energy $E$.
The reactant density of states $\rho^{\text{C}}_{\text{+}}(E)$
associated with crossing trajectories only (those trajectories that enter and exit
the region $\calM_+$ \cite{Berne82}) is then
\begin{equation}
\label{vol_1p}
\rho^{\text{C}}_{\text{+}}(E) = \phi(E) \, \sbar
\end{equation}
where the superscript $\text{C}$ indicates the restriction to crossing trajectories.
The result \eqref{vol_1p} is essentially the content of the so-called
classical spectral theorem
\cite{Brumer80,Pollak81,Binney85,Meyer86,WaalkensBurbanksWiggins05,WaalkensBurbanksWiggins05c}.

If \emph{all} points in the  region $\calM_+$  eventually leave that region (that is,
all points lie on crossing trajectories
\cite{DeLeon81,Berne82}) then
\begin{equation}
\label{equality_1}
\rho^{\text{C}}_{\text{+}}(E) = \rho_{\text{+}}(E),
\end{equation}
so that the crossing density of states is equal to
the full reactant phase space density of states.
Apart from a set of measure zero, all phase points $\bz \in \calM_{\text{+}}$
can be classified as either trapped (T) or crossing (C) \cite{Berne82}.
A phase point in the trapped region $\calM_{\text{+}}^{\text{T}}$ never crosses the DS(E),
so that the associated trajectory does not contribute to the reactive flux.
Phase points in the crossing region $\calM_{\text{+}}^{\text{C}}$ do however eventually
cross the dividing surface, and so lie on trajectories that contribute to the reactive flux.
In general, however, as a consequence of the existence of trapped trajectories
(either trajectories on invariant \emph{trapped} $2$-tori \cite{DeLeon81,Berne82} or
trajectories asymptotic to other invariant objects of zero measure),
we have the inequality \cite{Thiele62,Berne82,Hase83}
\begin{equation}
\label{vol_2}
\rho_{\text{+}}^{\text{C}}(E) \leq \rho_{\text{+}}(E).
\end{equation}

If $\rho_{\text{+}}^{\text{C}}(E) < \rho_{\text{+}}(E)$, then it is in principle necessary to
introduce corrections to statistical estimates of reaction rates
\cite{Berne82,Hase83,Gray87,Berblinger94,Grebenshchikov03,Stember07}.
Numerical computation of crossing and reactive densities of states
for the HCN molecule are discussed
in refs \onlinecite{WaalkensBurbanksWigginsb04,WaalkensBurbanksWiggins05,Ezra09a},
and results for the Hamiltonian isokinetic thermostat are discussed in ref.\ \onlinecite{collins10}.
Note that, if the strict inequality $\rho_{\text{+}}^{\text{C}}(E) < \rho_{\text{+}}(E)$ holds, then the
system dynamics \emph{cannot} be ergodic on the energy shell at energy $E$.
The equality $\rho_{\text{+}}^{\text{C}}(E) = \rho_{\text{+}}(E)$ is therefore a necessary
condition for ergodicity, one that can be checked numerically.

\subsection{Gap time and reactant lifetime distributions}
\label{subsec:gaps}

The \emph{gap time distribution}, $\calP(s; E)$ is of central interest in
unimolecular kinetics \cite{Slater56,Thiele62}: the probability
that a phase point on $\text{DS}_{\text{+}}(E)$ at energy $E$ has a gap time between
$s$ and $s +\rmd s$ is equal to $\calP(s; E) \rmd s$.
An important idealized gap distribution is the random, exponential distribution
\begin{equation}
\label{exp_1}
\calP(s; E) = k(E) \, e^{-k(E) s}
\end{equation}
characterized by a single decay constant $k$ (where $k$ depends on energy $E$),
with corresponding mean gap time $\sbar = k^{-1}$.
An exponential distribution of gap times is usually  taken to be
a necessary condition for `statistical' behavior
in unimolecular reactions \cite{Slater56,Slater59,Thiele62,Dumont86,Carpenter03}.

The lifetime (time to cross the dividing surface $\text{DS}_{\text{-}}(E)$)
of phase point $\bz(\widetilde{p}, \widetilde{A}, \psi)$ is $t = s(\widetilde{p}, \widetilde{A}) - \psi$, and
the corresponding  (normalized)
reactant lifetime distribution function $\bbP(t; E)$ at energy $E$ is
\cite{Slater56,Slater59,Thiele62,Bunker62,Bunker64,Bunker73,Dumont86}
\begin{subequations}
\label{life_1}
\begin{align}
\label{life_1a}
\bbP(t; E) &= -\frac{\rmd}{\rmd t'}\; \text{Prob}(t \geq t'; E) \Big\vert_{t'=t} \\
\label{life_1b}
&= \frac{1}{\sbar} \, \Int{s}{t}{+\infty} \calP(s; E)
\end{align}
\end{subequations}
where the fraction of interesting (reactive) phase points having lifetimes between $t$ and $t + \rmd t$ is
$\bbP(t; E) \rmd t$.  It is often useful to work with the unnormalized lifetime distribution $F$,
where $F(t; E) \equiv  \sbar \, \bbP(t; E)$.

Equation \eqref{life_1a} gives the general relation between the lifetime distribution and the
fraction of trajectories having lifetimes greater than a certain value for arbitrary ensembles
\cite{Bunker62,Bunker64,Bunker73}.
Note that an exponential gap distribution \eqref{exp_1}
implies that the reactant lifetime
distribution $\bbP(t; E)$ is also exponential
\cite{Slater56,Slater59,Thiele62,Bunker62,Bunker64,Bunker73}; both gap and lifetime distributions
for realistic molecular potentials have
been of great interest since the earliest days of trajectory simulations of
unimolecular decay, and many examples of non-exponential lifetime distributions
have been found
\cite{Thiele62a,Bunker62,Bunker64,Bunker66,Bunker68,Bunker73,Hase76,Grebenshchikov03,Lourderaj09}.

\subsection{Reaction rates and the inverse gap time}

The quantity
\begin{equation}
\label{k_RRKM}
k^{\text{RRKM}}_{f}(E) \equiv \frac{\phi(E)}{\rho_{\text{+}}(E)}
\end{equation}
is the statistical (RRKM) microcanonical rate for the forward reaction
(trajectories crossing $\text{DS}_{+}$) at energy $E$, the ratio of the magnitude of
the flux $\phi(E)$ through $\text{DS}_{\text{+}}(E)$
to the total reactant density of states \cite{Robinson72,Forst03}.

Clearly, if $\rho_{\text{+}}(E) = \rho_{\text{+}}^{\text{C}}(E)$, then
\begin{equation}
k^{\text{RRKM}}_{f}(E) = \frac{1}{\sbar}
\end{equation}
the inverse mean gap time.
In general, the inverse of the mean gap time is
\begin{subequations}
\label{k2}
\begin{align}
k & \equiv \frac{1}{\sbar} = \frac{\phi(E)}{\rho_{\text{+}}^{\text{C}}} \\ 
& = k^{\text{RRKM}}_f \, \left[\frac{\rho_{\text{+}}(E)}{\rho_{\text{+}}^{\text{C}}(E)}\right] \\
& \geq k^{\text{RRKM}}_f.
\end{align}
\end{subequations}
The inverse gap time can then be interpreted as the
statistical unimolecular reaction  rate corrected for the volume of trapped trajectories in the
reactant phase space \cite{Dumont86,Berne82,Hase83,Gray87,Berblinger94}.

\subsection{Reactive flux correlation function}
\label{subsec:reactive_flux}

The discussion of reactive fluxes across the phase space dividing surface separating reactant
from product and of gap times provides a theoretical framework for analyzing
the lifetime distribution of an ensemble of trajectories initiated in the reactant well
at constant energy, where the lifetime refers to the time to the first crossing of the
dividing surface.  

Another approach to isomerization kinetics considers an equilibrium (canonical or 
microcanonical) ensemble of reactants
and products.  The regression hypothesis relates the
total relaxation rate for an initial perturbation of the equilibrium
populations to the autocorrelation function of spontaneous population fluctuations \cite{Chandler87}.
Standard analysis \cite{Chandler87} then provides a relation between 
the isomerization rate, when the latter exists, 
and the computationally tractable quantity ${\cal K}(t)$ given in terms
of the reactive flux across the barrier:
\begin{equation}
\label{eq:rate_1}
{\cal K}(t) = 
\frac{1}{x_{+}x_{-}} \langle \dot{q}(0) \delta[q(0)-q^{\ddagger}] \Theta_{+}[q(t)]\rangle 
\sim \frac{e^{-t/\tau}}{\tau}.
\end{equation}
In this expression,  $q \equiv \Ab_1$, the reaction coordinate, and 
the DS is determined by symmetry, so that the critical value $q^{\ddagger} = 0$.
We also have
\begin{equation}
\frac{1}{\tau} = k_f + k_b = 2k_f.
\end{equation}
In our calculations the ensemble average $\langle \cdots \rangle$ corresponds to 
a microcanonical ensemble (average over the entire energy shell $\Sigma_E$), 
$\Theta_{+}$ is the characteristic function for the configuration space
region $\Ab_2 > 0$, 
and equilibrium fractions are $x_{+} = x_{-} = 1/2$.  In the limit $t \to 0$, the right hand side of
equation \eqref{eq:rate_1} is just twice the statistical
rate $k_{f}^{\text{RRKM}}$, eq.\ \eqref{k_RRKM}.

Operationally, in principle we must
sample the DS without regard to the sign of the initial
velocity $\dot{q}(0)$.  A trajectory contributes to the average \eqref{eq:rate_1}
at time $t$:
\begin{enumerate}

\item {Only} if the phase point is in the product well ($q>0$) at time $t$,

\item With a sign ($\pm$) determined by the \emph{initial} sign of $\dot{q}$.

\end{enumerate}

The right hand side of \eqref{eq:rate_1} decays to zero as $t \to \infty$, as trajectories initially
crossing from product to reactant ($\dot{q} < 0$) eventually return to the product side,
leading to cancellation.
If the right hand side of \eqref{eq:rate_1} exhibits a so-called `plateau' region in which it
is approximately constant, followed by exponential decay,
then an isomerization rate can be extracted from the computation. 
This behavior indicates a well-defined separation of timescales:  trajectories remain
trapped in either well for long times with only infrequent transitions (crossing of the DS)
between wells.
On the other hand, if the 
reactive flux correlation functions exhibits oscillatory decay, then no rate constant
exists at the energy in question.  Pioneering
computations of flux correlation functions
for a number of 2 DoF dynamical models for isomerization were made by DeLeon and Berne 
\cite{DeLeon81,Berne82}.

In practice, we exploit the symmetry of the potential, and sample only 
initial conditions with $\dot{\Ab}_1>0$.
If the fraction of the phase points in the product well $A_1>0$ at time $t$ is $W(t)$, then
the fraction of the phase points in the reactant well, $A_1<0$, at time $t$ is $1 - W(t)$. 
As the potential is symmetric about $A_1=0$, 
reversing the initial sign of $\dot{q}$ leads to a 
set of symmetry-related trajectories with the occupancy of the two wells 
now $1 - W(t)$ and $W(t)$, respectively.
Adding the contributions of trajectories to \eqref{eq:rate_1}
with appropriate sign yields a result proportional to $2 W(t) - 1$,
which can be calculated from $W(t)$ directly. 

A connection between the gap time and the reactive flux approaches to isomerization kinetics was 
established by Straub and Berne in their work on the ``absorbing boundary'' method for
computing isomerization rates \cite{Straub85,Straub85a}.  Assuming that  there are no 
correlations between successive crossings of the DS for a given trajectory (i.e., 
assuming ``chaotic'' dynamics), then the single-passage gap/lifetime distribution can be used
to derive an expression for the reactive flux, and hence the associated isomerization 
rate \cite{Straub85a}.

\newpage
\section{Parameter Values and Energy Scales}
\label{sec:param}

\subsection{Physical parameters}
\label{subsec:phys_param}

We study a 2-mode truncation of the dynamics of a 
silicon nanobeam having rectangular cross section under compressive stress,
subject to hinged boundary conditions. 
The following physical parameter values are used \cite{chakra07,chakra09,chakra11}:  
elastic modulus $Q = 1.3 \times 10^{11}$  J/m$^3$;
uncompressed length $L_0 = 5 \times 10^{-8}$ m; width $w = 2 \times 10^{-9}$ m; 
depth $d = 1 \times 10^{-9}$ m; density $\rho = 2330$ kg/m$^3$.
For these parameters, the critical value of the strain $\epsilon_c$, obtained by solving 
the implicit equation \eqref{eq:critical_strain}, is $\epsilon_c = -0.000329$.

\subsection{Strain values: 3 cases}
\label{subsec:strain}

For a beam described by the physical parameters listed above, we consider
the dynamics for 3 values of the compressive stress.  The corresponding
strain values and associated parameters are given in Table \ref{tab:cases}.
Strain values for the three cases are:
case I, $\epsilon = -0.00065840 \simeq 2 \times \epsilon_c$;
case II, $\epsilon = -0.00197520 \simeq 6 \times \epsilon_c$;
case III, $\epsilon = -0.001419692 \simeq 4 \times \epsilon_c$.
The strain value for case III is chosen so that
the energy of the index 2 saddle lies just above the pair of
index 1 saddles at $\beta = -0.0001$.

Contour plots of the potential \eqref{eq:pot1} for the 3 cases considered
are given in Figure \ref{fig:pot_plots}.

Setting coordinate $\Ab_2 =0$ in potential \eqref{eq:pot1}, we obtain a bistable 
potential which is a function of the `reaction coordinate' $\Ab_1$.
In Table \ref{tab:cases} we give the value of the barrier height $\Delta E$ for each
of the 3 cases (degrees K).  It can be seen that the 
barrier heights are comparable to thermal energies $\sim 100$ K for all cases.
We have also estimated the magnitude of vibrational quanta $\hbar \omega$ associated with 
oscillations of the beam along the reaction coordinate at the potential energy
minimum; these energies are given in Table \ref{tab:cases}.
We have $\hbar \omega/k_{\text{B}}T \ll 1$ for $T \gtrsim 100$ K.

\newpage

\section{Results and Discussion}
\label{sec:results}

\subsection{Reactive flux, phase space volumes and ergodicity}
\label{sec:ergodicity}

We have computed reactive fluxes associated with the symmetry-determined DS
$\Ab_2 = 0$ for each of the 3 cases listed in Table \ref{tab:cases} at
3 energies: $E = 10^{-9}$, $E=10^{-8}$ and $E=10^{-7}$.  
For cases II and III, where the coordinate origin $(0,0)$ is an index 2 saddle
on the potential energy surface flanked by a pair of index 1 saddles, 
we have also performed computations at several values of $E <0$.
For case II, where the energy of the index 2 saddle is well above the pair of
index 1 saddles, we have used 3 additional energies: $E = -2.12 \times
10^{-7}$, $E= -2 \times 10^{-7}$
and $E= -1 \times 10^{-7}$.
For case III where the energy of the index 2 saddle is only just above the pair of
index 1 saddles we consider 2 additional energies: $E = -4 \times 10^{-9}$
and $E= -2.5 \times 10^{-9}$.
In each case the lowest energy is close to the index 1 saddle energy.
Our numerical results, obtained via Monte Carlo sampling of the DS, are
presented in Tables \ref{tab:case_1}--\ref{tab:case_5}.
For further details on the numerical methods used in these 
computations, see ref.\ \onlinecite{collins10}.

Numerical results for mean gap time $\sbar$, reactive flux $\phi_{+}(E)$, 
reactant volume $\rho^{\text{C}}_{+}(E) = \sbar \times \phi_{+}$, 
reactant density of states $\rho_{+}(E)$, pulse decay constant $\kappa$ (see below)
and the statistical isomerization rate $k^{\text{RRKM}}_f$
for each case and energy studied are given in 
Tables \ref{tab:case_1}--\ref{tab:case_5}. 

For the simple quartic potential obtained by setting $\Ab_1 =0$, 
action integrals (fluxes) $I_2(E)$ for motion in the  invariant plane $\Pi_2$ 
can be computed explicitly as a function of total energy $E$
in terms of complete Elliptic integrals.
These analytical expressions, not reported here, have been used as a 
check on our numerical calculations.

Our results show that $\rho^{\text{C}}_{+}(E) < \rho_{+}(E)$ in all instances; in 
the majority of cases the phase space volume swept out by reacting (crossing) trajectories is
considerably smaller than the full classical density of states associated with the
reactant region of phase space.  This means that, for the stress values and energies studied here,
the buckled nanobeam dynamics is very far from being ergodic.
Ergodicity is usually taken to be a necessary (but by no means sufficient)
condition for the applicability of statistical theories of reaction rates.

Some representative trajectories for the 3 cases are shown in Figure 
\ref{fig:case_2_traj}.  
 For each case/energy we present two plots: one  shows 20 trajectories initiated on the DS
 and followed until the first recrossing of the DS, while the other shows
 a single trajectory followed for 200 crossings of the DS.  
 It is clear by inspection of the single trajectory plots that the dynamics is far from ergodic 
 on the timescale considered; the trajectories appear to be quasiperiodic or weakly chaotic at most.

\subsection{Gap time distribution}
\label{sec:gaptime}

Both gap time distributions $\calP (t)$ and associated (unnormalized) lifetime  
distributions $F(t)$ have been computed for all cases.
The functions $\calP(t)$ and $\log[F(t)]$ are plotted in Figure \ref{fig:gap_time_plots}
for the lowest energy in each  case.
The results shown represent the range of behavior 
found for the various cases and energies we have studied.

For case I, $E=10^{-7}$, 
the gap time distribution essentially consists of a single `pulse' 
associated with 
trajectories that enter the well, exhibit a single turning point in the reaction
coordinate, and then exit through the dividing surface DS$_{-}$.
The smallest gap time is nonzero, reflecting
a delay corresponding to the time it takes for a point on the shortest lived trajectory
to reach the turning point and then return to the DS.
The lifetime distribution
of the set of trajectories in the pulse is however well described by a single exponential 
decay, with a decay rate (denoted $\kappa$) that is 
much faster than either the RRKM rate $k^{\text{RRKM}}_f$ 
or the inverse of the mean gap time.

For cases II and III, the structure of the gap time distribution 
is more complex, consisting of 
multiple pulses.  Several early pulses have comparable amplitudes, 
with amplitude not necessarily decreasing monotonically with gap time.
In such cases the computation of decay rates for
individual pulses is less accurate due to the possibility of overlapping pulses.
At long gap times the distribution becames smooth, with typically nonexponential decay.

Representative pulse decay constants $\kappa$ are listed in Tables \ref{tab:case_1}--\ref{tab:case_5}. 
In all instances $\kappa \gg k^{\text{RRKM}}_f$.

We note that the gap time distributions seen here are reminiscent of the
`epistrophic' patterns of ionization times seen in the work of Mitchell and Delos \cite{Delos06}.

\subsection{Reactive flux correlation function}
\label{sec:fluxcorr}

The quantity ${\cal K}(t)$ (eq.\ \eqref{eq:rate_1}) 
is plotted in Figure \ref{fig:flux_plots}
for three different cases/energies.  
The behavior seen in these plots is again typical of all the 
cases we have examined:  ${\cal K}(t)$ exhibits
oscillatory decay over a much longer timescale than the gap time
decay constant $\kappa$.  The absence  of a
`plateau' region means that an isomerization rate
constant \emph{does not exist} for our
nanobeam model in the physical regime studied.

\subsection{Gap time distribution on the DS}
\label{sec:TSgap}

To further explore the isomerization dynamics of the 
nanobeam, we examine the distribution of gap times on
the dividing surface.  That is, we plot contours of
the gap time $s$ as a function of coordinates $(\Ab_2, p_2)$ 
on DS$_{+}$.

A set of representative plot is shown in Figure \ref{fig:DS_gap_time}, corresponding
to case I, $E= 10^{-7}$, case II, $E= 10^{-9}$ and case III, $E= 10^{-9}$.  
Note that the contour plots are
invariant with respect to
the inversion operation $(\Ab_2, p_2) \to (-\Ab_2, - p_2)$.
For each case the gap time $s$ is also plotted along the line $p_2 = 0$.

A significant finding is that for case I the gap time is a smooth function 
of location on the DS; although there appear to be no singularities inside the 
boundary (NHIM) with divergent gap times, closer examination (not shown here) confirms  the
existence of a
small region near the NHIM (the boundary of the DS) 
where gap times are longer, presumably associated with second (and later) pulses
that do not cross the TS on
first approach but bounce back again one (or more) times.
The absence of `fractal' patterns such as those seen in previous 
studies \cite{WaalkensBurbanksWigginsb04} indicates that the intra-well
dynamics is extremely simple for case I. 

The gap time contours  on the DS for cases II and III show a 
more typical \cite{WaalkensBurbanksWigginsb04} fractal arrangement.
Initiial condition on the DS for which the gap time diverges
presumably lie on the stable manifold of either the NHIM or of
a periodic orbit confined to the reactant well.
By symmetry, trajectories on the line $p_2=0$ having divergent gap times
must lie on both the stable manifold (in forward time) and unstable 
manifold (in backward time) of the NHIM, and hence lie on homoclinic orbits.

\newpage
\section{Conclusions and outlook}
\label{sec:conc}

In this paper we have studied the classical (Euler-Bernoulli) mechanics  
of a 2-mode truncation 
of the  dynamics for a buckled nanobeam with rectangular cross section 
subject to compressive stress.  The physical parameters used correspond to 
a Silicon nanobeam \cite{lawrence02,carr05,lawrence06,chakra07,chakra09,chakra11}.  In the stress      
regime studied, the first transverse displacement mode has become unstable, 
while the second mode is either stable 
or unstable, depending on the value of the strain.
The resulting beam Hamiltonian has the same form as model 2 DoF systems
previously studied in chemistry, which describe a bistable reaction (isomerization)
coordinate coupled to an additional transverse or `bath' mode.

We have applied methods from reaction rate theory 
to characterize the nanobeam `isomerization' dynamics.  For the beam model considered, the 
dividing surface separating `reactant' and `product'  configurations for the buckled
beam is completely determined by symmetry (coordinate $\Ab_1 = 0$).  
Using exponential dichotomies, we have proved that, 
for a specified range of the energy, the 
boundary of the associated dividing surface is a Normally Hyperbolic Invariant Manifold (NHIM).

We have computed reactive fluxes, mean gap times and reactant phase space volumes
for 5 stress values at several different energies.
In all cases the phase space volume swept out by crossing trajectories is considerably
less that the reactant density of states, proving that the dynamics
is highly nonergodic.  The associated gap time distributions consist 
of single `pulses' of trajectories.  Computation of the reactive flux 
correlation function shows no
sign of a plateau region; rather, the flux exhibits oscillatory decay, indicating that,
for the 2-mode model in the physical regime considered, a rate constant for
isomerization does not exist.  

Problems for future work include study of the dynamical influence of additional `bath' modes, and
the investigation of quantum effects at low temperatures.


\acknowledgments

PC and SW  acknowledge the support of the  Office of Naval Research (Grant No.~N00014-01-1-0769) and the Leverhulme Trust.

\newpage

\appendix

\section{Proof of the Existence of a Normally Hyperbolic Invariant Manifold}
\label{sec:nhim_exist}

Roughly speaking, a normally hyperbolic invariant manifold has the property that, under the dynamics linearized about the invariant manifold, growth rates in directions
transverse to the invariant manifold dominate the growth rates of directions tangent to the invariant manifold.  (For some background on normally hyperbolic invariant
manifolds see refs \onlinecite{fenichel71, fenichel74, fenichel77, hps77}.
An account of Fenichel's approach to the theory, as well as  some history and examples can be found in
ref.\ \onlinecite{Wiggins94}. )
The dynamics on $\Pi_2$ are completely integrable. For $\beta>0$ it consists entirely of periodic orbits, and the growth rates (e.g. Lyapunov
exponents)  associated with all orbits are all zero.
For $\beta <0$ it consists entirely of
periodic orbits, {\em except} for the saddle point at the origin connected by a pair of homoclinic
orbits.  The periodic orbits all have zero growth rates, and the saddle point and the homoclinic orbits will be discussed separately. We will show that the growth rates
transverse to
$\Pi_2$ are exponential. Hence they  dominate the growth rates tangent to $\Pi_2$.

We linearize \eqref{eq:hameq_1} about $\Pi_2$ and evaluate the resulting equations on an arbitrary trajectory on $\Pi_2$.
Since $\Pi_2$ is a plane, and is described in a global coordinate system,
linearization about $\Pi_2$ is particularly easy.  Coordinates describing the  directions normal to
$\Pi_2$ are $(\Ab_1, \pb_1)$, and  $\Pi_2$  is defined by   $\Ab_1=\pb_1=0$ in the four dimensional phase space
with coordinates  $(\Ab_1, \pb_1, \Ab_2, \pb_2)$.

Therefore, the linearized dynamics about $\Pi_2$ is obtained by retaining terms only linear in $(\Ab_1, \pb_1)$:
\begin{subequations}
\begin{align}
\label{eq:hameq_lin}
\dot{\Ab}_1 & =  \pb_1,  \\
\dot{\pb}_1 & =  - 2 \Ab_1 \left( \alpha + 4\Ab_2^2  \right),  \\
\dot{\Ab}_2 & =  \pb_2,  \\
\dot{\pb}_2 & =   - 8 \Ab_2 \left( \beta + 4\Ab_2^2 \right).
\end{align}
\end{subequations}
The linearized dynamics normal to $\Pi_2$ are given by:
\begin{subequations}
\label{eq:hameq_lin_1}
\begin{align}
\dot{\Ab}_1 & =  \pb_1,  \\
\dot{\pb}_1 & =  - 2 \Ab_1 \left( \alpha + 4\Ab_2^2(t)  \right).
\end{align}
\end{subequations}
where $\Ab_2^2(t) $ is a component of a trajectory on $\Pi_2$, i.e. it is  the $\Ab_2$  component of a trajectory of:
\begin{subequations}
\begin{align}
\label{eq:hameq_lin_2}
\dot{\Ab}_2 & =  \pb_2, \\
\dot{\pb}_2 & =  - 8 \Ab_2 \left( \beta + 4\Ab_2^2 \right).
\end{align}
\end{subequations}

We now will show that the linear, nonautonomous  equation \eqref{eq:hameq_lin_1} has  two linearly independent solutions;
one exponentially growing in time, and the
other exponentially decaying in time.  This is accomplished by showing that \eqref{eq:hameq_lin_1}
has an {\em exponential dichotomy}, and these conditions will depend
on $\alpha, \, \beta, \, \Ab_2^2(t)$.

The numbers that are typically used to quantify (linearized) growth rates of trajectories are the Lyapunov exponents. However, for nonautonomous ordinary differential
equations exponential dichotomies are often more convenient and facilitate the proof of certain results. A discussion of the relationship between Lyapunov exponents and
exponential dichotomies can be found in, e.g., refs \onlinecite{dieci1, dieci3}.

\paragraph{Exponential Dichotomy}

First, we will provide some background on the notion of exponential dichotomy that is
particular to our situation, i.e. two dimensional time-dependent ordinary differential
equations. The standard reference on exponential dichotomies is ref.\ \onlinecite{coppel}.

Consider the linear ordinary differential
equation with time dependent coefficients
\begin{equation}
\dot{\mathbf{x}} = L(t) \mathbf{x},
\label{nalin}
\end{equation}
where $\mathbf{x}=(x,y)$ and the $2 \times 2$ matrix $L(t)$ is a continuous
function of $t$, and suppose $X(t)$ is the fundamental solution matrix
of (\ref{nalin}).  Let $\parallel \cdot \parallel$ denote a matrix norm,
such as the maximum of the absolute values of the matrix elements.
Then (\ref{nalin}) is said to possess an {\em exponential dichotomy}
if there exists	a rank-one projection operator $P$ and constants
$K_1, \, K_2, \, \alpha_1, \,\alpha_2 > 0$, such	that
\begin{subequations}
\begin{align}
\parallel X(t) P X^{-1}(\tau) \parallel	& \le
K_1 \exp \left(- \alpha_1 (t - \tau) \right),
\quad	t \ge \tau,  \label{ineq1} \\
\parallel X(t) \left( \mbox{id}	-  P \right)  X^{-1}(\tau) \parallel & \le
K_2 \exp \left(\alpha_2	(t - \tau) \right), \quad t \le	\tau.  \label{ineq2}
\end{align}
\end{subequations}

The condition that the  projection operator $P$ has rank one means that,
of the two linearly independent solutions of (\ref{nalin}), one is
exponentially growing and one is exponentially decaying.

Verifying that \eqref{eq:hameq_lin_1} has an exponential dichotomy requires us to solve for the  fundamental solution matrix $X(t)$. In general, this is not possible.
Instead, we will use  results on ''roughness of exponential dichotomies'' \cite{coppel, jw}.  The relevant result is as follows.  Suppose \eqref{nalin} has the form:
\begin{equation}
\dot{\mathbf{x}} = \left( A(t) + B(t) \right)\mathbf{x},
\label{nalin_2}
\end{equation}
and suppose that the equation:
\begin{equation}
\dot{\mathbf{x}} =  A(t) \mathbf{x},
\label{nalin_3}
\end{equation}
has an exponential dichotomy with constants $K_1, \, K_2, \, \alpha_1, \, \alpha_2$. If
\begin{equation}
\sup_{t \in  \mathbb{R}} \parallel B(t) \parallel  \left( \frac{K_1}{\alpha_1} + \frac{K_2}{\alpha_2} \right) < 1,
\label{cond_1}
\end{equation}
then \eqref{nalin_2} has an exponential dichotomy with constants $K_1' = K_2' = K_3>0$ and $\alpha_1' = \alpha_2' = \alpha_3>0$.

We now apply these ideas to \eqref{eq:hameq_lin_1}.  Rewriting this equation in the form  of \eqref{nalin} gives:
\begin{equation}
\left(
\begin{array}{c}
\dot{\Ab}_1  \\
\dot{\pb}_1
\end{array}
\right) = \left[
\left(
\begin{array}{cc}
0 & 1 \\
-2 \alpha & 0
\end{array}
\right) +
\left(
\begin{array}{cc}
0 & 0 \\
-8  \Ab_2^2 (t) & 0
\end{array}
\right)
\right]
\left(
\begin{array}{c}
\Ab_1  \\
\pb_1
\end{array}
\right)
\label{eq:ed1}
\end{equation}
We  introduce a linear change of coordinates that diagonalizes the first matrix in this expression:
\begin{equation}
\left(
\begin{array}{c}
\Ab_1  \\
\pb_1
\end{array}
\right)  = T
\left(
\begin{array}{c}
x \\
y
\end{array}
\right)
\end{equation}
where
\begin{equation}
T = \left(
\begin{array}{cc}
1 & 1 \\
\sqrt{-2 \alpha} & -\sqrt{-2 \alpha}
\end{array}
\right), \quad T^{-1} =\frac{-1}{ 2 \sqrt{-2 \alpha}}
\left(
\begin{array}{cc}
 -\sqrt{-2 \alpha} & -1 \\
-\sqrt{-2 \alpha} & 1
\end{array}
\right)
\end{equation}
Then \eqref{eq:ed1} becomes:
\begin{equation}
\left(
\begin{array}{c}
\dot{x}  \\
\dot{y}
\end{array}
\right) = \left[
\left(
\begin{array}{cc}
\sqrt{-2 \alpha}  & 1 \\
0 & -\sqrt{-2 \alpha}
\end{array}
\right) -\frac{1}{\sqrt{-2 \alpha}}
\left(
\begin{array}{cc}
4\Ab_2^2 (t) & 4\Ab_2^2 (t) \\
- 4\Ab_2^2 (t) & -4\Ab_2^2 (t)
\end{array}
\right)
\right]
\left(
\begin{array}{c}
x  \\
y
\end{array}
\right)
\label{eq:ed2}
\end{equation}

First we consider:
\begin{equation}
\left(
\begin{array}{c}
\dot{x}  \\
\dot{y}
\end{array}
\right) =
\left(
\begin{array}{cc}
\sqrt{-2 \alpha}  & 1 \\
0 & -\sqrt{-2 \alpha}
\end{array}
\right) \left(
\begin{array}{c}
x  \\
y
\end{array}
\right)
\label{eq:ed3}
\end{equation}
The fundamental solution matrix is given by:
\begin{equation}
X(t) = \left(
\begin{array}{cc}
e^{\sqrt{-2 \alpha} \, t }& 0 \\
0 & e^{-\sqrt{-2 \alpha} \, t }
\end{array}
\right)
\label{fsm}
\end{equation}

We take as the projection matrix:
\begin{equation}
P = \left(
\begin{array}{cc}
0 & 0 \\
0 & 1
\end{array}
\right)
\label{proj}
\end{equation}
Then, using \eqref{proj}  and \eqref{fsm}, we have:
\begin{subequations}
\begin{equation}
 X(t) P X^{-1}(\tau) 	= \left(
 \begin{array}{cc}
 0 & 0 \\
 0 & e^{-\sqrt{-2 \alpha} \, (t-\tau)}
 \end{array}
 \right)
\end{equation}
and
\begin{equation}
X(t) \left( \mbox{id}	-  P \right)  X^{-1}(\tau) = \left(
 \begin{array}{cc}
 e^{\sqrt{-2 \alpha} \, (t-\tau)} & 0 \\
 0 & 0
 \end{array}
 \right)
\end{equation}
\end{subequations}

It follows that \eqref{eq:ed3} has an exponential dichotomy with $K_1 = K_2=1$ and $\alpha_1 = \alpha_2 = \sqrt{-2 \alpha}$.
We now need to check \eqref{cond_1}. For \eqref{eq:ed2} this condition takes the form:
\begin{equation}
\frac{4 \,\sup_{t \in \mathbb{R}} \vert \Ab_2^2 (t) \vert}{- \alpha} <1.
\label{cond_2}
\end{equation}

The quantity $\sup_{t \in \mathbb{R}} \vert \Ab_2^2 (t) \vert$ is the maximum value that $\Ab_2^2 (t)$ attains along a trajectory, 
and this can be  precisely
computed, as we now show.

In Section \ref{sec:nhim} we showed the the dynamics in the $\Ab_2-\pb_2$ plane is Hamiltonian, with Hamiltonian given by:
\begin{equation}
\label{eq:ham_2}
\Hb_2 = \frac{\pb_2^2}{2} + 4\beta \Ab_2^2 + 8 \Ab_2^4.
\end{equation}
Trajectories lie on level curves of the Hamiltonian $\Hb_2$:
\begin{equation}
 \frac{\pb_2^2}{2} + 4\beta \Ab_2^2 + 8 \Ab_2^4= E,
 \label{eq:levelset}
\end{equation}
where
\begin{equation}
\begin{array}{cc}
E \ge 0 & \quad \mbox{for} \quad \beta \ge 0, \\
E > -\frac{1}{2} \beta^2 & \quad \mbox{for} \quad \beta <0.
\end{array}
\end{equation}

The quantity $\sup_{t \in \mathbb{R}} \vert \Ab_2^2 (t) \vert$ corresponds to the largest
``turning point''  of a trajectory, i.e., the largest value of $\Ab_2^2$ that
intersects the $\Ab_2$ axis.  An equation for this quantity can be obtained by setting $\pb_2=0$ in \eqref{eq:levelset}.
Doing so, and rearranging terms, gives the
following quadratic equation for $\Ab_2^2$:
\begin{equation}
\Ab_2^4 +  \frac{\beta}{2} \Ab_2^2 - \frac{E}{8}=0.
\end{equation}
The solution of this  equation is:
\begin{equation}
\Ab_2^2 = -\frac{\beta}{4} \pm \frac{1}{4}\sqrt{\beta^2 + 2E}
\end{equation}
and we will take the plus sign in front of the square root since
we are seeking the largest root. It is also useful to  note that this expression is an increasing function of
$E$ since:
\begin{equation}
\frac{d \Ab_2^2 }{dE} = \frac{1}{4 \,\sqrt{\beta^2 + 2E}} >0.
\end{equation}

Therefore we seek the largest value of $E$ such that \eqref{cond_2} is satisfied.
This is obtained by equating the expression for
$\Ab_2^2 = -\frac{\beta}{4} + \frac{1}{4}\sqrt{\beta^2 + 2E}$ to $-\frac{\alpha}{4}$, which gives:
\begin{equation}
 -\beta +\sqrt{\beta^2 + 2E} = -\alpha.
 \end{equation}
 and then solving this expression for $E$:
 \begin{equation}
 E_{\text max} = \frac{\alpha^2}{2} \left(1-2 \frac{\beta}{\alpha} \right).
 \end{equation}

\def\cprime{$'$} \def\cprime{$'$}


\newpage

\begin{table}[H]
\begin{center}
\begin{tabular}{|c|c|c|} \hline
 Equilibrium point & Eigenvalues  & Energy of the equilibrium \\ \hline\hline
$\left(\bar{A}_1, \bar{p}_1, \bar{A}_2, \bar{p}_2 \right) = \left(0, 0, 0, 0 \right)$
 &  $\pm \sqrt{-2 \alpha}, \, \pm 2 \sqrt{-2 \beta}$ & $0$  \\
 \hline
$\left(\bar{A}_1, \bar{p}_1, \bar{A}_2, \bar{p}_2 \right) = \left(\pm \sqrt{-\alpha}, 0, 0, 0 \right)$
 & $\pm 2\sqrt{\alpha}, \, \pm 2 \sqrt{2(\alpha - \beta)}$ & $-\frac{1}{2} \alpha^2$   \\
 \hline
$\left(\bar{A}_1, \bar{p}_1, \bar{A}_2, \bar{p}_2 \right) = \left(0, 0, \pm\frac{\sqrt{-\beta}}{2}, 0 \right)$
 & $\pm 4 \sqrt{\beta}, \, \pm \sqrt{2(\beta - \alpha)}$ & $-\frac{1}{2} \beta^2 $ \\
 \hline
\end{tabular}
\caption{\label{table:eqinfo} The location of the equilibria, the eigenvalues of the matrix associated with the linearization of Hamilton's equations about the equilibria, and
the (total) energy of the equilibrium. In the last two rows both equilbrium points 
have the same four eigenvalues.  }
\end{center}
\end{table}

\vspace*{1.cm}

\begin{table}[H]
\begin{center}
\begin{tabular}{|c|c|c|c|c|c|c|} \hline
Case & $\epsilon$  & $\bar{\epsilon}$  &  $\alpha$  & $\beta$ & 
$\Delta E/ k_{\text{B}}$ [K] & $\hbar \omega/k_{\text{B}}$ [K] \\      \hline \hline
I & -0.00065840 & -0.00032942 & -0.00032898 & 0.00065928 & 50.98 & 0.092 \\      \hline
II & -0.00197520 & -0.00033029 & -0.00164491 & -0.00065404 &  1274.4 &  0.206  \\      \hline
III & -0.00141969 & -0.00032992 & -0.00108977 & -0.00010000 &  559.4  & 0.168 \\      \hline
\hline
\end{tabular}
  \caption{\label{tab:cases} Strain values $\epsilon$ and associated 
  parameter values $\bar{\epsilon}$, $\alpha$ and
  $\beta$ for the 3 cases used in our computations. Also shown are barrier heights $\Delta E$ and
  estimates of the size of vibrational quanta for beam oscillations about the energy minimum
  (degrees K).}
\end{center}
\end{table}

\vspace*{1.cm}

\begin{table}[htpb]
\begin{center}
\begin{tabular}{|c|c|c|c|c|c|c|c|} \hline
Energy & $\sbar$ & $\phi_{\text{+}}(E)$ & $\rho_{\text{+}}^{\text{C}}(E)$ & $\rho_{\text{+}}(E)$ &
$k = \sbar^{-1}$ & $k^{\text{RRKM}}_f$ & $\kappa$ \\ \hline
\hline 1e-09 & 357.460 & 0.00000009 &  0.000031 & 0.00085 & 0.00280 & 0.0001013 & 0.0254\\
\hline 1e-08 & 266.663 & 0.00000086 &  0.000230 & 0.00103 & 0.00375 & 0.0008351 & 0.0248\\
\hline 1e-07 & 174.982 & 0.00000832 &  0.001456 & 0.00210 & 0.00571 & 0.0039613 & 0.0224\\
\hline
\end{tabular}
\caption{\label{tab:case_1}
Computational results for case I: $\alpha =  -0.00032898$, $\beta = 0.00065928$.  For discussion
see Sec.\ \ref{sec:results}.}
\end{center}
\end{table}

\begin{table}[htpb]
\begin{center}
\begin{tabular}{|c|c|c|c|c|c|c|c|} \hline
Energy & $\sbar$ & $\phi_{\text{+}}(E)$ & $\rho_{\text{+}}^{\text{C}}(E)$ & $\rho_{\text{+}}(E)$ &
$k = \sbar^{-1}$ & $k^{\text{RRKM}}_f$ & $\kappa$ \\ \hline
\hline -2.12e-07 & 7729.396 & 0.00000023 &  0.001789 & 0.00897 & 0.00013 & 0.0000258 & 0.0645\\
\hline -2e-07 & 1258.603 & 0.00000172 &  0.002160 & 0.00914 & 0.00079 & 0.0001878 & 0.0406\\
\hline -1e-07 & 385.348 & 0.00001488 &  0.005734 & 0.01034 & 0.00260 & 0.0014386 & 0.0481\\
\hline 1e-09 & 166.588 & 0.00003186 &  0.005308 & 0.01137 & 0.00600 & 0.0028028 & 0.0312\\
\hline 1e-08 & 165.124 & 0.00003384 &  0.005588 & 0.01132 & 0.00606 & 0.0029886 & 0.0454\\
\hline 1e-07 & 141.782 & 0.00004771 &  0.006764 & 0.01183 & 0.00705 & 0.0040325 & 0.0492\\
\hline
\end{tabular}
\caption{\label{tab:case_2}
Computational results for case II: $\alpha =  -0.00164491$, $\beta = -0.00065404$.
For discussion see Sec.\ \ref{sec:results}.}
\end{center}
\end{table}

\begin{table}[htpb]
\begin{center}
\begin{tabular}{|c|c|c|c|c|c|c|c|} \hline
Energy & $\sbar$ & $\phi_{\text{+}}(E)$ & $\rho_{\text{+}}^{\text{C}}(E)$ & $\rho_{\text{+}}(E)$ &
$k = \sbar^{-1}$ & $k^{\text{RRKM}}_f$ & $\kappa$ \\ \hline
\hline -4e-09 & 4076.711 & 0.00000032 &  0.001308 & 0.00574 & 0.00025 & 0.0000559 & 0.0618\\
\hline -2.5e-09 & 1567.930 & 0.00000083 &  0.001304 & 0.00579 & 0.00064 & 0.0001436 & 0.0427\\
\hline 1e-09 & 502.955 & 0.00000236 &  0.001187 & 0.00590 & 0.00199 & 0.0004001 & 0.0402\\
\hline 1e-08 & 307.775 & 0.00000497 &  0.001528 & 0.00600 & 0.00325 & 0.0008280 & 0.0621\\
\hline 1e-07 & 195.281 & 0.00001941 &  0.003791 & 0.00680 & 0.00512 & 0.0028533 & 0.0410\\
\hline
\end{tabular}
\caption{\label{tab:case_5}
Computational results for case III: $\alpha =  -0.00108977$, $\beta = -0.0001$. 
For discussion see Sec.\ \ref{sec:results}.}
\end{center}
\end{table}


\clearpage

\section*{Figure captions}

\begin{figure}[H]
\caption{Phase space portraits in the invariant $\Ab_2-\pb_2$ plane.  (a)  $\beta >0$ (b) $\beta <0$.
}
\label{fig:nhim}
\end{figure}

  \begin{figure}[H]
  \caption{Contour plots of the 2-mode nanobeam potential
  eq.\ \eqref{eq:pot_dim} for the 3 compressive stress values
  considered in this paper. The contour values shown include the particular energies 
  at which the dynamics was studied.
  (a) $\alpha =  -0.00032898 , \,  \beta = 0.00065928$;
  (b) $\alpha =  -0.00164491, \,  \beta = -0.00065404$;
  (c) $\alpha =  -0.00108977, \,  \beta = -0.0001$.
  }
  \label{fig:pot_plots}
  \end{figure}

    \begin{figure}[H]
    \caption{Plots of trajectories initiated on the DS.
      (a), (b) Case I, energy $E=10^{-7}$; 
      (c), (d) case II, energy $E= 10^{-9}$;
      (e), (f) case III, energy $E=10^{-9}$.
      Panels (a), (c) and (e) each show 20 trajectories followed until the
      first recrossing of the DS, while 
      panels (b), (d) and (f) show single trajectories 
      followed for 200 crossings of the DS.
      }
    \label{fig:case_2_traj}
    \end{figure}

\begin{figure}[H]
\caption{Gap time distribution $\calP (t)$ and the logarithm of the associated 
(unnormalized) lifetime distribution $F(t)$.
  (a), (b) case I, energy $E= 10^{-7}$;
  (c), (d) case II, energy $E=10^{-9}$;
  (e), (f) case III, energy $E=10^{-9}$.
}
\label{fig:gap_time_plots}
\end{figure}

\begin{figure}[H]
\caption{Reactive flux correlation function ${\cal K}(t)$ versus $t$.
(a) case I, energy $E=10^{-7}$;
(b) case II, energy $E=10^{-9}$;
(c) case III, energy $E=10^{-9}$.
}
\label{fig:flux_plots}
\end{figure}

\begin{figure}[H]
\caption{Contours of the gap time $s$ as a function of coordinates 
$(\Ab_2, p_2)$ on the dividing surface DS$_{+}$ (panels (a), (c), (e))
and as a function of $\Ab_2$ along the line $p_2 = 0$ (panels (b), (d) and (e)).
(a), (b) case I, energy $E=10^{-7}$;
(c), (d) case II, energy $E=10^{-9}$;
(e), (f) case III, energy $E=10^{-9}$.
}
\label{fig:DS_gap_time}
  \end{figure}


\newpage

\begin{center}
 \includegraphics[width=14cm]{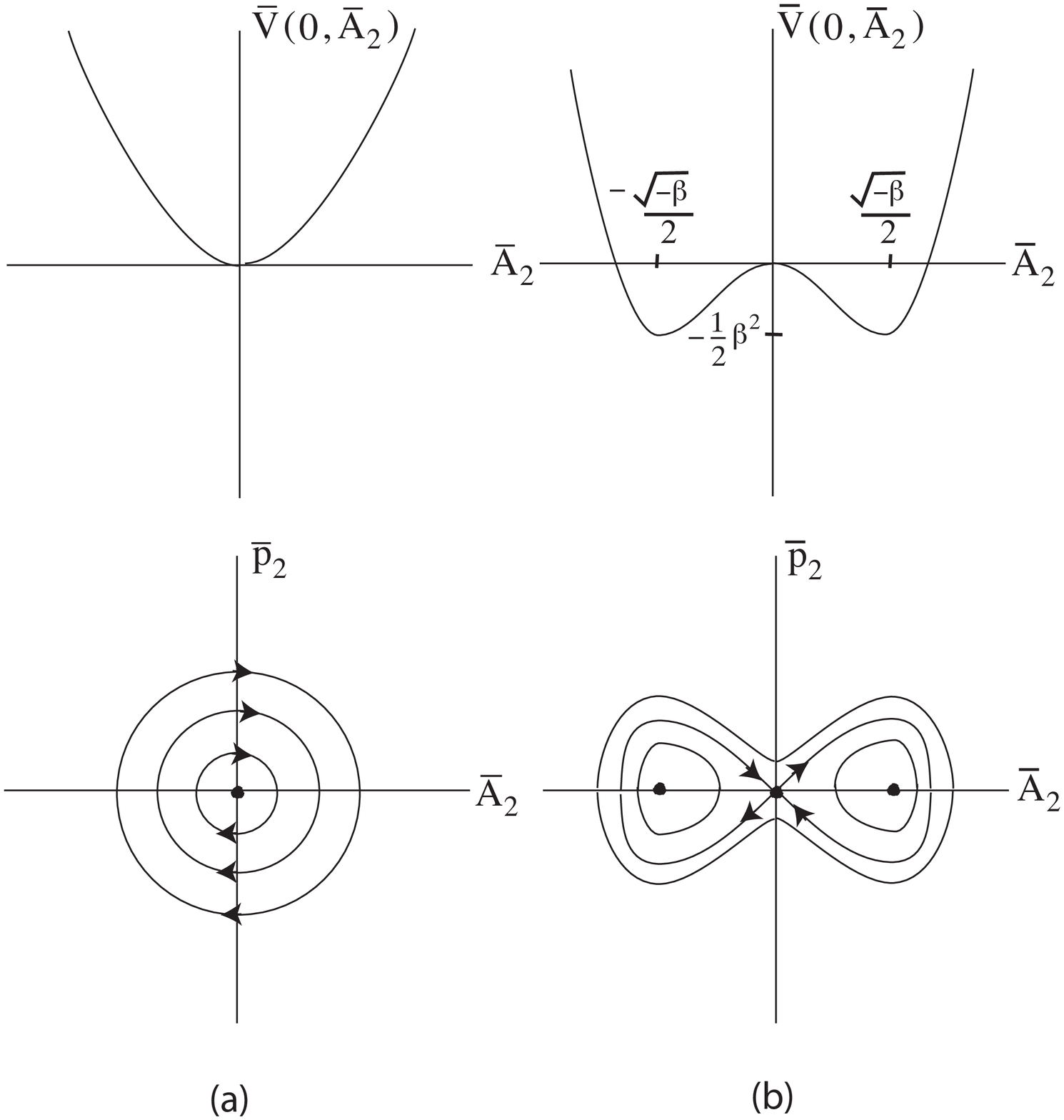}
\end{center}

 \vspace*{1.5cm}
 FIGURE 1

\newpage

\begin{center}
 \includegraphics[width=18cm]{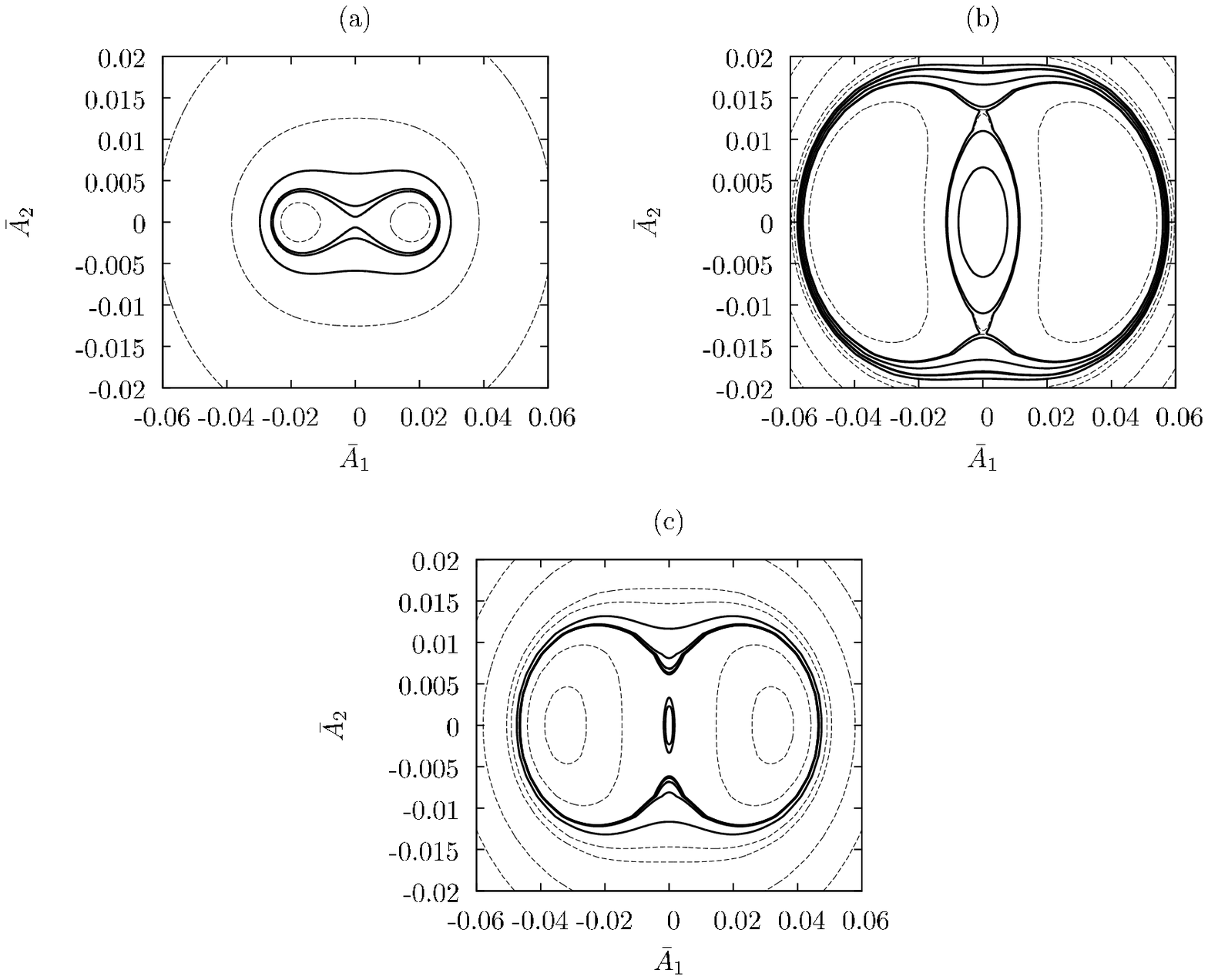}
\end{center}

 \vspace*{1.5cm}
 FIGURE 2

\newpage

\begin{center}
 \includegraphics[width=14cm]{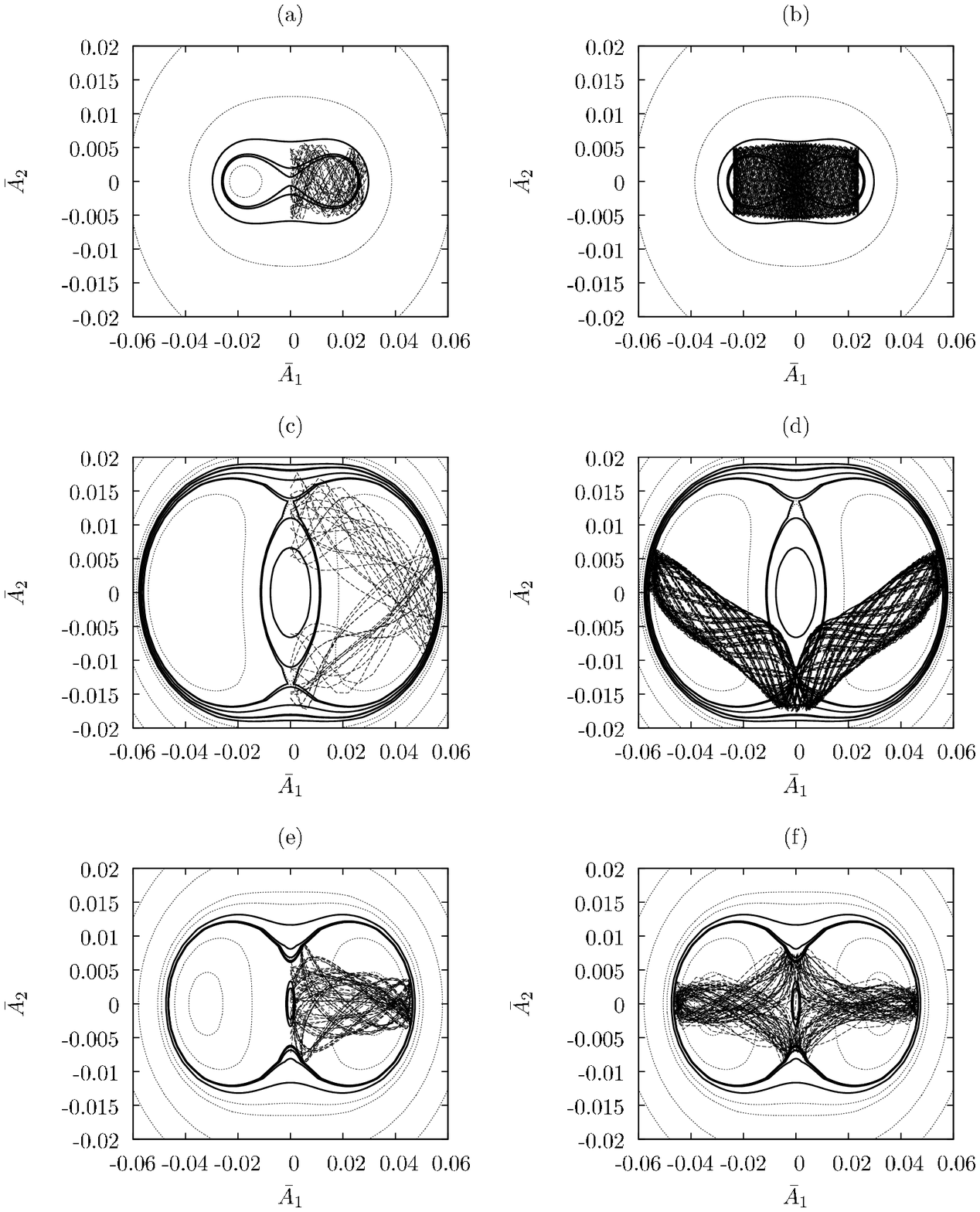}
\end{center}

 \vspace*{1.5cm}
 FIGURE 3

\newpage

\begin{center}
 \includegraphics[width=14cm]{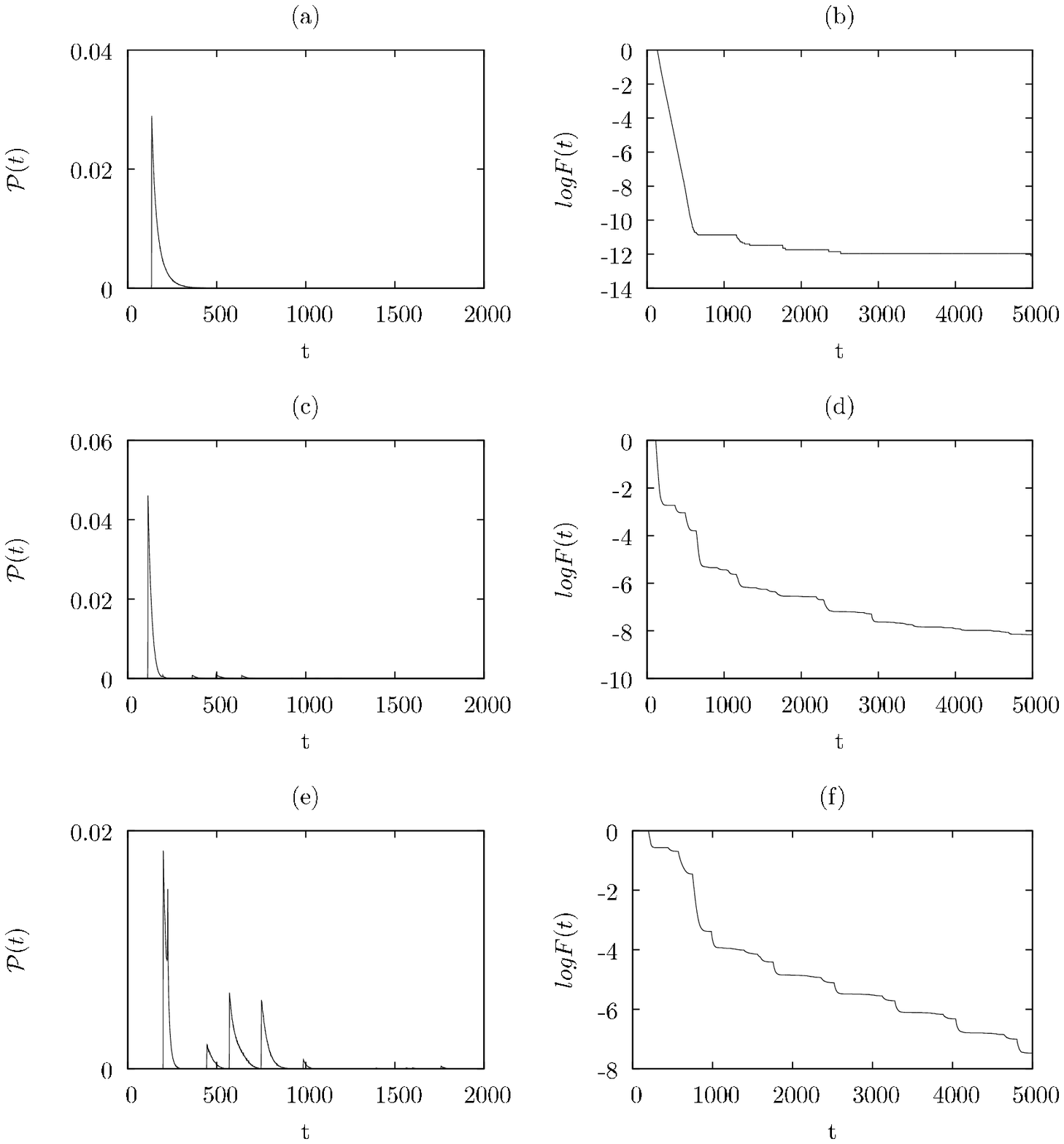}
\end{center}

 \vspace*{1.5cm}
 FIGURE 4

 \newpage
 
 \begin{center}
  \includegraphics[width=8cm]{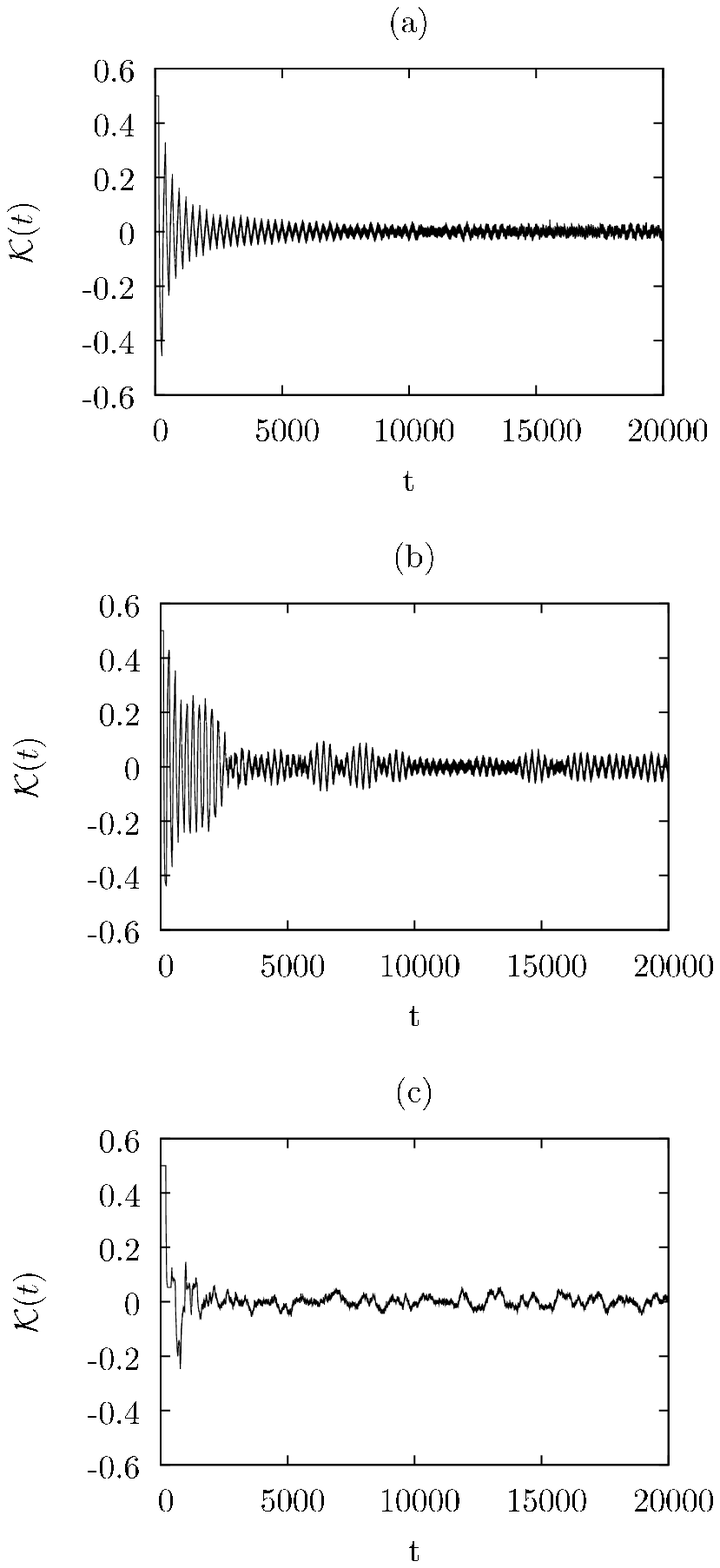}
 \end{center}
 
  \vspace*{1.5cm}
  FIGURE 5

 \begin{center}
  \includegraphics[width=14cm]{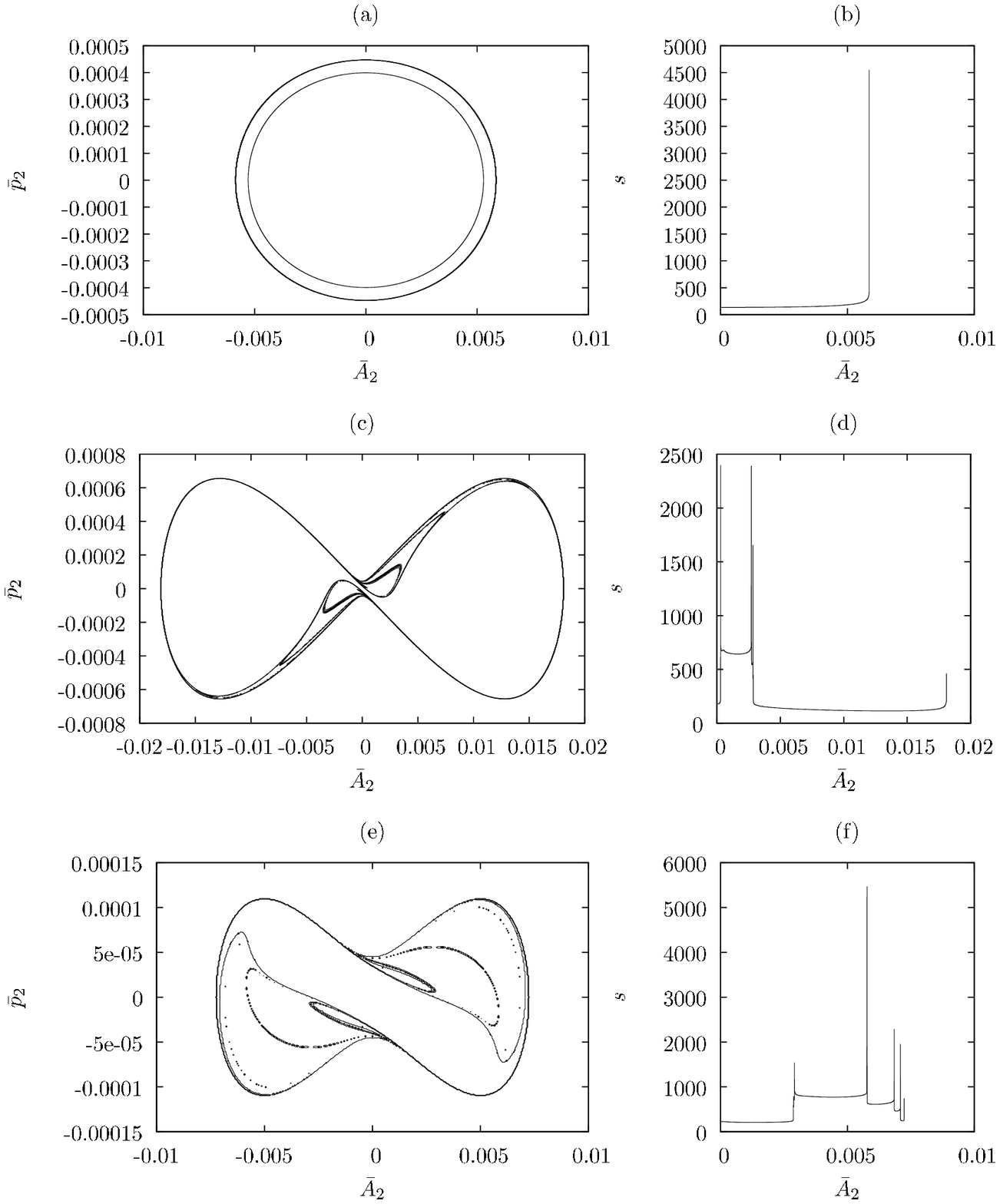}
 \end{center}
 
  \vspace*{1.5cm}
  FIGURE 6

\end{document}